# Transcriptome signature for the identification of bevacizumab responders in ovarian cancer

Olga Zolotareva[1*], Karen Legler[2], Olga Tsoy[1], Anna Esteve[3], Alexey Sergushichev[4], Vladimir Sukhov[4], Jan Baumbach[1,5], Kathrin Eylmann[2], Minyue Qi[6], Malik Alawi[6], Stefan Kommoss[7,8], Barbara Schmalfeldt[2], Leticia Oliveira-Ferrer[2]

[1]Institute for Computational Systems Biology, University of Hamburg, Hamburg, Germany

[2] Department of Gynecology, University Medical Center Hamburg-Eppendorf, Hamburg, Germany

[3]Medical Oncology Service, Catalan Institute of Oncology, Campus Can Ruti; Badalona-Applied Research Group in Oncology (B-ARGO), Germans Trias I Pujol Research Institute (IGTP); Research Management Unit (UGR), Catalan Institute of Oncology.

[4]Washington University in St. Louis School of Medicine, Department of Pathology and Immunology

[5]Computational BioMedicine lab, University of Southern Denmark, Odense, Denmark

[6]Bioinformatics Core, University Medical Center Hamburg-Eppendorf, Hamburg, Germany

[7]Department of Gynecology and Obstetrics, University Hospital Tuebingen, Calwerstrasse 7, 72076 Tuebingen, Germany.

[8]Department of Gynecology and Obstetrics, Diak Klinikum Schwaebisch Hall, Diakoniestrasse 10, 74523 Schwaebisch Hall, Germany.

*Corresponding author:

Olga Zolotareva

Institute for Computational Systems Biology

University of Hamburg

E-mail: olga.zolotareva@uni-hamburg.de

# Abstract

The standard of care for ovarian cancer comprises cytoreductive surgery, followed by adjuvant platinum-based chemotherapy plus taxane therapy and maintenance therapy with the antiangiogenic compound bevacizumab and/or a PARP inhibitor. Nevertheless, there is currently no clear clinical indication for the use of bevacizumab, highlighting the urgent need for biomarkers to assess the response to bevacizumab.

In the present study, based on a novel RNA-seq dataset (n=181) and a previously published microarray-based dataset (n=377), we have identified an expression signature potentially associated with benefit from bevacizumab addition and assumed to reflect cancer stemness acquisition driven by activation of CTCFL. Patients with this signature demonstrated improved overall survival when bevacizumab was added to standard chemotherapy in both novel (HR=0.41(0.23-0.74), adj.p-value=7.70e-03) and previously published cohorts (HR=0.51(0.34-0.75), adj.p-value=3.25e-03), while no significant differences in survival explained by treatment were observed in patients negative for this signature.

In addition to the CTCFL signature, we found several other reproducible expression signatures which may also represent biomarker candidates not related to established molecular subtypes of ovarian cancer and require further validation studies based on additional RNA-seq data.

# Introduction

Ovarian cancer is the second leading cause of death from gynecological cancer with an overall 5-year survival rate of less than 50%. This high mortality is mainly due to the lack of effective screening tools, leading to 75% of women being diagnosed with advanced stages and a high recurrence rate[1]. Standard of care consists of cytoreductive surgery followed by adjuvant platinum-based plus taxane therapy and maintenance therapy with the antiangiogenic compound bevacizumab and/or a PARP inhibitor[2].

PARP inhibitors have shown significant efficacy, especially in ovarian cancers associated with BRCA mutations or homologous recombination deficiency. PARP inhibitors exploit the concept of synthetic lethality, where the combination of two genetic or pharmacological defects leads to cell death[3]. Bevacizumab is an antibody against vascular endothelial growth factor (VEGF), a key factor in the regulation of angiogenesis. In cancer, VEGF inhibition slows or prevents the growth of blood vessels that supply the tumor with oxygen and nutrients, leading to tumor necrosis. Several studies have shown that the addition of bevacizumab during first-line therapy, followed by 10 months of monotherapy with bevacizumab, leads to a significant extension of progression-free survival (PFS) by 4 months [4–6]. It is important to point out that the use of bevacizumab in cancer treatment leads to potential side effects, such as hypertension and gastrointestinal bleeding/perforation, and the decision to include it in a treatment plan is usually based on an assessment of the potential benefits versus risks for the patient. In this context, the identification of patients who will benefit from antiangiogenic therapy is crucial, and many efforts have been made over the last decade to identify biomarkers of response to bevacizumab in ovarian cancer.

In the context of the ICON7 trial, several serum biomarkers have been tested: a discriminatory signature comprising mesothelin, FLT4, AGP, and CA-125 has been described to potentially identify those patients with epithelial ovarian cancer (EOC) more likely to benefit from the antiangiogenic therapy[7], and also the combined values of serum Ang1 and Tie2 levels has been proposed as predictive biomarkers for improved PFS in bevacizumab-treated patients with ovarian cancer[8]. Additionally, serum VEGF-A or VEGF-C levels have been identified in other cohorts to be significantly associated with response to bevacizumab therapy[9,10]. Also, YKL-40, a proangiogenic glycoprotein secreted by cancer and inflammatory cells as a consequence of VEGF inhibition, is associated with improved outcomes in patients with chemotherapy-refractory advanced ovarian cancer treated with bevacizumab[11].

In the context of the GOG-0218 clinical trial[12], five candidate tumor biomarkers (VEGF-A, CD31, neuropilin-1, MET, VEGFR-2) were assessed by immunohistochemistry in an extensive population (n = 980), but only higher microvessel density (measured by CD31) showed predictive value for PFS. Tumor VEGF-A was not predictive for PFS but showed potential predictive value for overall survival (OS) using a third-quartile cutoff for high VEGF-A expression[13]. In contrast, VEGF-A165b, a VEGF-A splice variant, was analyzed in a tissue microarray of the ICON7 trial, suggesting that bevacizumab may improve the prognosis of patients with advanced ovarian cancer with low expression of this VEGFA isoform[14]. In the same cohort, immunofluorescence staining for co-localised c-MET and VEGFR-2 and genotyping of germline DNA from peripheral blood leukocytes for VEGFA and VEGFR-2 SNPs were evaluated. Here, high c-MET/VEGFR-2 co-localisation on tumor tissue and the VEGFR-2 rs2305945 G/G variant, which may be biologically related, were associated with worse survival outcomes in bevacizumab-treated women diagnosed with EOC[15]. In addition, Volk et al. showed that patients with high Ang-2 expression benefit significantly from therapy with bevacizumab and that the predictive role of Ang-2 is limited to its expression in the tumor compartment, implying that Ang-2 could serve as a tissue-based biomarker but is not suitable as a blood-based biomarker[16].

At the gene expression level, data from the ICON7 and the MITO16A-ManGo-OV2 phase IV trial have been evaluated using DASL gene expression arrays and qRT-PCR, respectively, for the identification of signatures of bevacizumab response[17,18]. Kommoss et al. found that ovarian carcinoma molecular subtypes with the poorest survival (proliferative and mesenchymal) show a higher benefit from treatment including bevacizumab. Califano et al. found high miR-484 expression to be associated with longer progression-free and overall survival. Interestingly, the combined expression of miR-484 and its target VEGFB identified a subset of patients that might mostly benefit from bevacizumab treatment.

Most of the biomarkers mentioned above, both individual markers and signatures, have not yet been validated in independent cohorts, which lowers their relevance for clinical application. To enable personalization of bevacizumab therapy in EOC and identify reproducible biomarker candidates, in the present study, we generated a novel transcriptome dataset, the UKE cohort, comprising 244 ovarian tumor samples obtained from 212 patients treated at our clinic. Complete clinical information was available for 181 patients, of whom 114 were treated only with platinum-based therapy and 67 additionally received bevacizumab as maintenance therapy. We utilized unsupervised and supervised machine learning techniques to discover candidate expression signatures with potential predictive significance in bevacizumab-treated EOC. A previously published EOC cohort DASL[17] with similar patient characteristics and treatment settings was used as a validation dataset. Additionally, the existence of the candidate expression signature discovered in the UKE and DASL data, and its relationship with known molecular subtypes of EOC, was confirmed using another publicly available RNA-seq-based dataset, TCGA-OV.

# Results and discussion

## The UKE ovarian cancer RNA-Seq cohort

|  |  | Bevacizumab (n=67) | Standard (n=114) | Total (n=181) |
|---|---|---|---|---|
| **FIGO** |  |  |  |  |
|  | I | 0 | 2 (1.8%) | 2 (1.1%) |
|  | II | 0 | 5 (4.4%) | 5 (2.8%) |
|  | III | 54 (80.6%) | 83 (72.8%) | 137 (75.7%) |
|  | IV | 13 (19.4%) | 24 (21.1%) | 37 (20.4%) |
| **Grading** |  |  |  |  |
|  | low grade (G1/G2) | 4 (6.0%) | 19 (16.7%) | 23 (12.7%) |
|  | high grade | 61 (91.0%) | 95 (83.3%) | 156 (86.2%) |
|  | unknown | 2 (3.0%) | 0 | 2 (1.1%) |
| **Stage** |  |  |  |  |
|  | pT1-pT2 | 5 (7.5%) | 6 (5.3%) | 11 (6.1%) |
|  | pT3a-pT3b | 9 (13.4%) | 7 (6.1%) | 16 (8.8%) |
|  | pT3c | 51 (76.1%) | 100 (87.7%) | 151 (83.4%) |
|  | unknown | 2 (3.0%) | 1 (0.9%) | 3 (1.7%) |
| **pN** |  |  |  |  |
|  | N0 | 12 (17.9%) | 24 (21.1%) | 36 (19.9 %) |
|  | N1 | 43 (64.2%) | 65 (57.0%) | 108 (59.7 %) |
|  | Nx | 12 (17.9%) | 25 (21.9%) | 37 (20.4 %) |
| **Histological Type** |  |  |  |  |
|  | serous | 62 (92.5 %) | 107 (93.9 %) | 169 (93.4 %) |
|  | other or unknown | 5 (7.5 %) | 7 (6.1 %) | 12 (6.6 %) |
| **Surgery outcome** |  |  |  |  |
|  | no macroscopically visible tumor | 39 (58.2%) | 73 (64.0%) | 112 (61.9%) |
|  | tumor ≤ 1cm | 14 (20.9%) | 27 (23.7%) | 41 (22.7%) |
|  | tumor ≥ 1cm | 14 (20.9%) | 14 (12.3%) | 28 (15.5%) |
| **Age at diagnosis** | years, mean (CI95%) | 59.4 (56.4-62.4) | 60.3 (58.3-62.2) | 59.9 (58.3 - 61.6) |
| **Median follow-up** | months, median | 41.0 | 41.5 | 41.0 |
| **Events at median follow-up** | death | 35 (52.2%) | 85 (74.6%) | 120 (66.3%) |
|  | recurrence | 55 (82.1%) | 93 (81.6%) | 148 (81.8%) |

**Table 1.** The baseline characteristics of the UKE cohort.

Most of the clinical characteristics of the UKE cohort summarized in Table 1 are similar to the characteristics of the DASL cohort previously published by Kommoss et al., 2017[17]. In both cohorts, patients were treated with either platinum-based chemotherapy (standard) or with platinum-based chemotherapy and subsequent maintenance therapy with bevacizumab (bevacizumab). While the Kommoss study includes 189 and 170 patients in the bevacizumab and standard treatment groups, our cohort comprised 67 and 114 patients, who received bevacizumab and standard respectively. The OS of these two treatment patient subgroups is comparable in both cohorts (Fig. 1a and 1b). However, bevacizumab demonstrates a pronounced favorable impact on PFS in the DASL cohort, while in the UKE cohort prolongation of PSF under bevacizumab treatment is not statistically significant (Fig. 1c and 1d). This discrepancy may be attributed to the different distribution and number of patients in the treatment groups. In particular, the bevacizumab group in the UKE cohort included patients with more advanced tumors (FIGO stage III or IV) compared to the DASL cohort where 13.8% were at an early stage (FIGO stage I or II)[17].

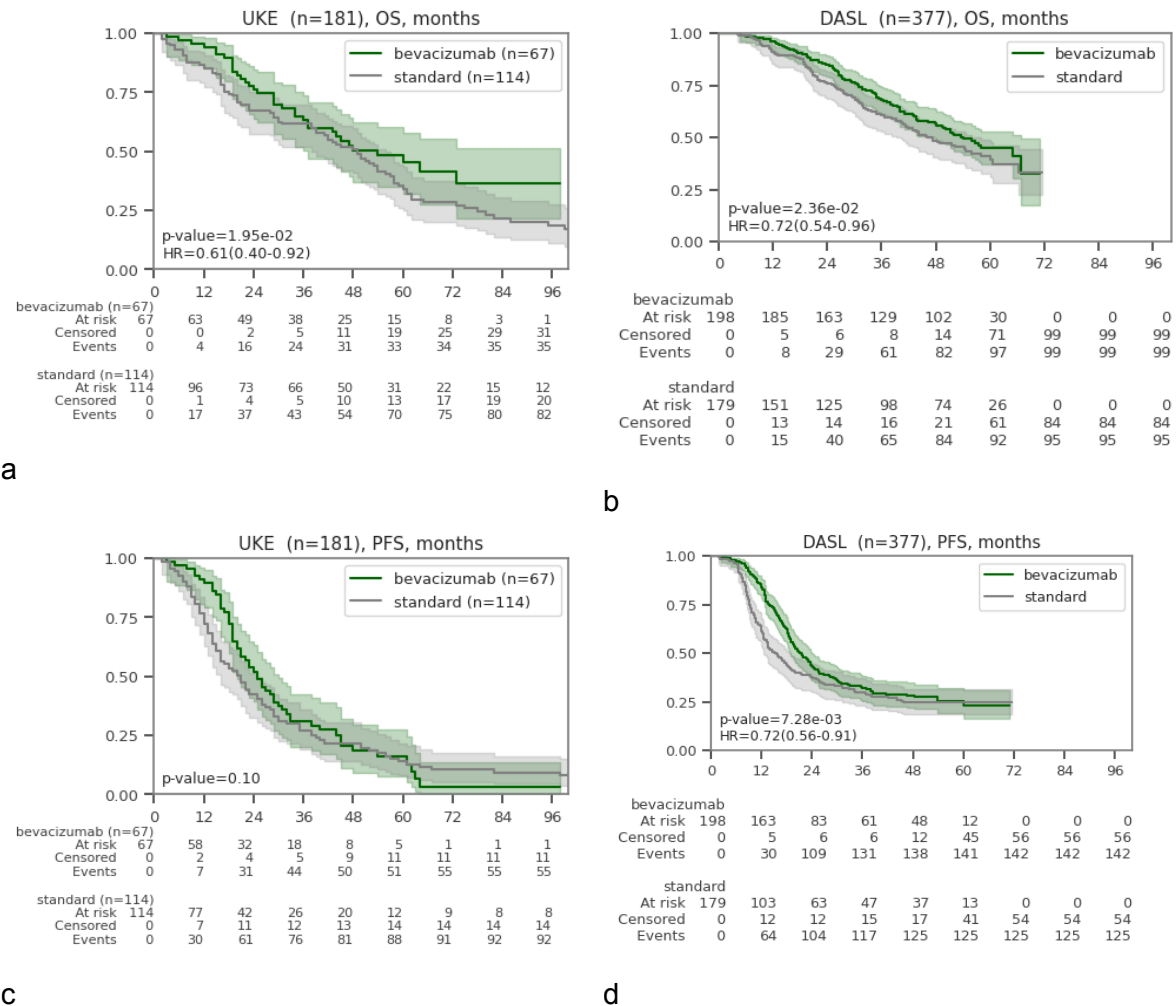

**Figure 1.** Kaplan-Meier plots for OS (**a,b**) and PFS (**c,d**) in bevacizumab and standard treatment groups of the UKE (**a,c**) and DASL (**b,d**) cohorts. X-axes show time from diagnosis in months. In the UKE cohort, bevacizumab addition to standard treatment significantly improved OS (HR=0.61; p-value = 0.02), but the improvement was not statistically significant for PFS (HR=0.75; p-value = 0.1). In the DASL cohort, the addition of bevacizumab to standard treatment significantly prolonged both OS (HR=0.72; p-value = 0.02), and PFS (HR=0.72; p-value = 0.007).

## Predictive value of individual genes

Since the expressions of certain genes were recently shown to be predictive for bevacizumab response in ovarian cancer[7–9,12,16], we evaluated predictive values of individual gene expressions in the UKE cohort. For this purpose, we employed the Cox Proportional Hazards models, which incorporated the patient's age, tumor stage, and surgery outcome as covariates (expression of each gene was tested independently; see Methods for details). After the Benjamini-Hochberg procedure, no gene expressions significantly predictive (adjusted p-value <0.05) for OS or PFS in the UKE or DASL cohorts (Supplementary Figure S1).

The lack of statistically significant single-gene biomarkers may be explained by the limited size of the UKE and DASL cohorts compared to the number of tested genes, and the associations between expressions and survival might be neglected after correction for multiple testing. Therefore, to mitigate the multiple testing problem, we performed supervised and unsupervised stratification of the UKE cohort based on the expression data, identified transcriptionally distinct sample sets and their specific biomarkers, and evaluated their predictive power in both cohorts.

## Predictive value of known molecular subtypes

Molecular subtypes were determined using a random forest-based classifier implemented in *consensusOV*[19] R package (Supplementary Table S2 and Methods). This consensus classifier trained on tumors from multiple independent datasets, concordantly subtyped by three previously published classifiers developed by Konecny et al.[20], Verhaak et al.[21], and Helland et al.[22], and demonstrated improved robustness compared to individual classifiers. Out of 181 samples in the UKE cohort, 27 were classified as proliferative, 56 as mesenchymal, 48 as differentiated, and 50 as immunoreactive.

Although patients who received bevacizumab in addition to standard treatment demonstrated improved OS, no molecular subtype was significantly predictive for OS or PFS in the UKE cohort (Table 2). Recently, *Kommoss et al.* suggested that molecular subtypes of ovarian cancer may demonstrate differential response to bevacizumab. They have shown that in the DASL cohort, patients with proliferative tumors may have improved PFS if bevacizumab is added to standard treatment[17]. Our analysis of the UKE cohort stratified by subtype also did not confirm this association, possibly due to limited cohort size (Supplementary Fig. S2). Other reasons for this discrepancy may be the smaller number and higher malignancy of patients in the UKE cohort or the difference in the approaches to the classification of patients into four molecular subtypes. Although molecular subtypes were shown to be prognostic for survival in many previous studies[19–21,23,24], in the UKE cohort, neither for patients under standard treatment only (Supplementary Fig. S3), nor for those who received bevacizumab (Supplementary Fig. S4), tumor subtypes demonstrated statistically significant association with OS or PFS in stratified analysis.

|  | bevacizumab | standard | OS | | PFS | |
|---|---|---|---|---|---|---|
|  |  |  | HR (95%CI) | p-value | HR (95%CI) | p-value |
| **Proliferative** | 13 | 14 | 0.73 (0.24-2.25) | 0.59 | 1.58 (0.60-4.17) | 0.35 |
| **Mesenchymal** | 23 | 33 | 1.14 (0.49-2.65) | 0.75 | 0.53 (0.26-1.10) | 0.09 |
| **Differentiated** | 7 | 41 | 0.56 (0.12-2.54) | 0.45 | 1.11 (0.41-2.98) | 0.83 |
| **Immunoreactive** | 24 | 26 | 1.46 (0.61-3.25) | 0.39 | 1.35 (0.64-2.84) | 0.43 |

**Table 2.** Predictive value of molecular subtypes in the UKE cohort. Interaction p-values are not adjusted for multiple testing.

## Unsupervised patient stratification identifies reproducible expression signatures

Since no statistically significant associations between four ovarian cancer subtypes and bevacizumab response in the UKE cohort were found, we further searched for other expression-based predictors unrelated to known molecular subtypes. For that we employed UnPaSt[25], an unsupervised patient stratification method searching for differentially expressed biclusters in omics data. A differentially expressed bicluster in transcriptome data represents a submatrix stratifying all samples into two subpopulations with high or low expression of a certain gene set (see Methods for details). Some biclusters correspond to molecular pathways whose activity may define molecular subtypes or correlate with clinical characteristics of tumors, including drug response.

In total UnPaSt identified 234 differentially expressed biclusters consisting of 2-128 genes and 11-90 samples in the UKE dataset (Supplementary Table S2). However, due to the platform differences, many genes of these biclusters were absent in the DASL validation data. To focus the search on reproducible biomarker candidates, we first evaluated whether the genes comprising each bicluster discovered in the UKE data could also stratify samples into high- and low-expression groups in the DASL dataset. Consequently, 23 biclusters including at least two genes and presenting in both UKE and DASL cohorts were considered for further analysis.

## Random forest prioritizes predictive bicluster candidates

The 23 biclusters detectable in both UKE and DASL datasets were encoded as binary variables and used to train random survival forest models together with the patient age, operation result and FIGO stage. For each of the two treatment groups, an independent model was trained for OS and PFS on the UKE using five-fold cross-validation and validated on the DASL data. All four models achieved the performance of 0.6 on the validation data, suggesting that they could only partially explain the observed survival. Feature importance analysis revealed that the strongest survival predictors were different for two treatment groups (Table 3). In the bevacizumab-treated group, patient age at diagnosis was the strongest predictor of both OS and PFS. In contrast, in the standard treatment group the importance of age was not high, and residual and tumor size > 1cm and FIGO stage IV had the highest importance.

|  | OS | | PFS | |
|---|---|---|---|---|
|  | **bevacizumab** | **standard** | **bevacizumab** | **standard** |
| **Performance** | | | | |
| DASL (validation) | 0.60 | 0.60 | 0.60 | 0.60 |
| UKE (training) | 0.75 | 0.69 | 0.65 | 0.68 |
| **Selected features** | | | | |
| Age at diagnosis | **0.040** | 0.009 | **0.032** | 0.004 |
| Bicluster 70 | **0.021** | -0.001 | **0.010** | 0.001 |
| Bicluster 84 | **0.013** | 0 | -0.004 | 0 |
| Bicluster 130 | **0.010** | 0 | 0.003 | 0.005 |
| Bicluster 90 | 0.005 | 0 | 0 | -0.001 |
| Bicluster 109 | 0.001 | 0 | 0.002 | 0 |
| Bicluster 41 | 0 | -0.001 | **0.016** | 0.003 |
| Bicluster 60 | 0 | 0.009 | 0 | 0.001 |
| Bicluster 95 | 0 | 0.003 | 0.009 | 0 |
| Bicluster 72 | 0 | 0 | 0 | 0.001 |
| Bicluster 144 | -0.004 | 0 | -0.002 | 0.004 |
| residual tumor > 1cm | 0.001 | **0.045** | 0.006 | **0.026** |
| residual tumor < 1cm | 0 | 0.009 | **0.021** | **0.012** |
| FIGO IV | 0 | **0.014** | 0 | **0.039** |
| FIGO IIIC | 0 | 0 | 0.004 | 0 |

**Table 3.** The results of training random survival forest models to predict OS and PFS in treatment groups. The top two rows report model performance on the validation and training data, measured as the concordance index (c-index). The next rows provide the strongest predictors of OS and PFS specific to each treatment group selected based on their permutation-based importances. Permutation-based feature importance is calculated as the average decrease of random survival forest model performance when the corresponding feature column in the validation dataset was permuted. Importance values exceeding 0.01 are shown in bold text font.

Out of 23 tested biclusters, only five had positive importance for OS and five for PFS in the bevacizumab-treated group, and only bicluster 70 had relatively high importance for both OS and PFS in the bevacizumab-treated group. Kaplan-meier plots for patients groups stratified by these bicluster candidates can be found in the Supplementary Figures S5 and S6.

Further, the predictive value of each bicluster candidate prioritized by random survival forest was evaluated for statistical significance using a Cox model adjusted for patient's age, tumor stage, and surgery outcome. Bicluster 84 represented the most promising predictive biomarker candidate for the OS under bevacizumab treatment compared to the standard treatment in both cohorts (Table 4).

|  | UKE (n=181) | | | DASL (n=377) | | |
| --- | --- | --- | --- | --- | --- | --- |
|  | p-value | adj.p-value | HR (CI 95%) | p-value | adj.p-value | HR (CI 95%) |
| **Bicluster 84** | 0.069179 | 0.23 | 2.1 (0.94-4.66) | 0.013626 | 0.07 | 2.06 (1.16-3.66) |
| **Bicluster 130** | 0.28696 | 0.36 | 1.59 (0.68-3.74) | 0.134611 | 0.34 | 1.77 (0.84-3.72) |
| **Bicluster 70** | 0.090687 | 0.23 | 0.42 (0.16-1.14) | 0.513614 | 0.80 | 1.21 (0.68-2.16) |
| **Bicluster 90** | 0.183871 | 0.31 | 1.75 (0.77-4.01) | 0.924283 | 0.92 | 1.03 (0.58-1.82) |
| **Bicluster 109** | 0.533283 | 0.53 | 1.31 (0.56-3.02) | 0.64217 | 0.80 | 1.17 (0.61-2.23) |

**Table 4.** Interaction between treatment and bicluster candidates predictive for OS selected by random survival forests. Each bicluster candidate was tested independently, the resulting p-values were adjusted following the Benjamini-Hochberg procedure.

In the DASL cohort, bicluster 84 signature significantly interacted with treatment (p-val.=1.36e-02) and demonstrated the same tendency in the UKE cohort, although it did not pass the commonly used threshold for statistical significance (p-val.=6.92e-02). In both cohorts, patients with overexpression of bicluster 84 genes in the tumor tissue benefited from the addition of bevacizumab to standard treatment, whereas bevacizumab treatment of patients with low expression of bicluster 84 did not provide any benefits compared to standard platinum-based therapy (Figure 3a-b). In the subset of the UKE cohort including 99 patients whose tumors over-expressed bicluster 84 genes, bevacizumab-treated group demonstrated a prolonged OS, compared to those who only received standard treatment (p-value=3.08e-03; adj.p-value=7.70e-03; HR=0.41(0.23-0.74). In the DASL cohort, bevacizumab addition caused the same effect on the OS of patients who over-expressed this signature (n=195;p-value =6.50e-04; adj.p-value=3.25e-03; HR=0.51(0.34-0.75). In both cohorts, no significant differences in the OS between treatment groups were observed among the patients whose tumors under-expressed the bicluster 84 genes (p-value=0.97, HR = 1.01(0.56-1.84), and p-value=0.87;HR =1.04(0.67-1.60), in the UKE and DASL datasets respectively;Figure 3c-d).

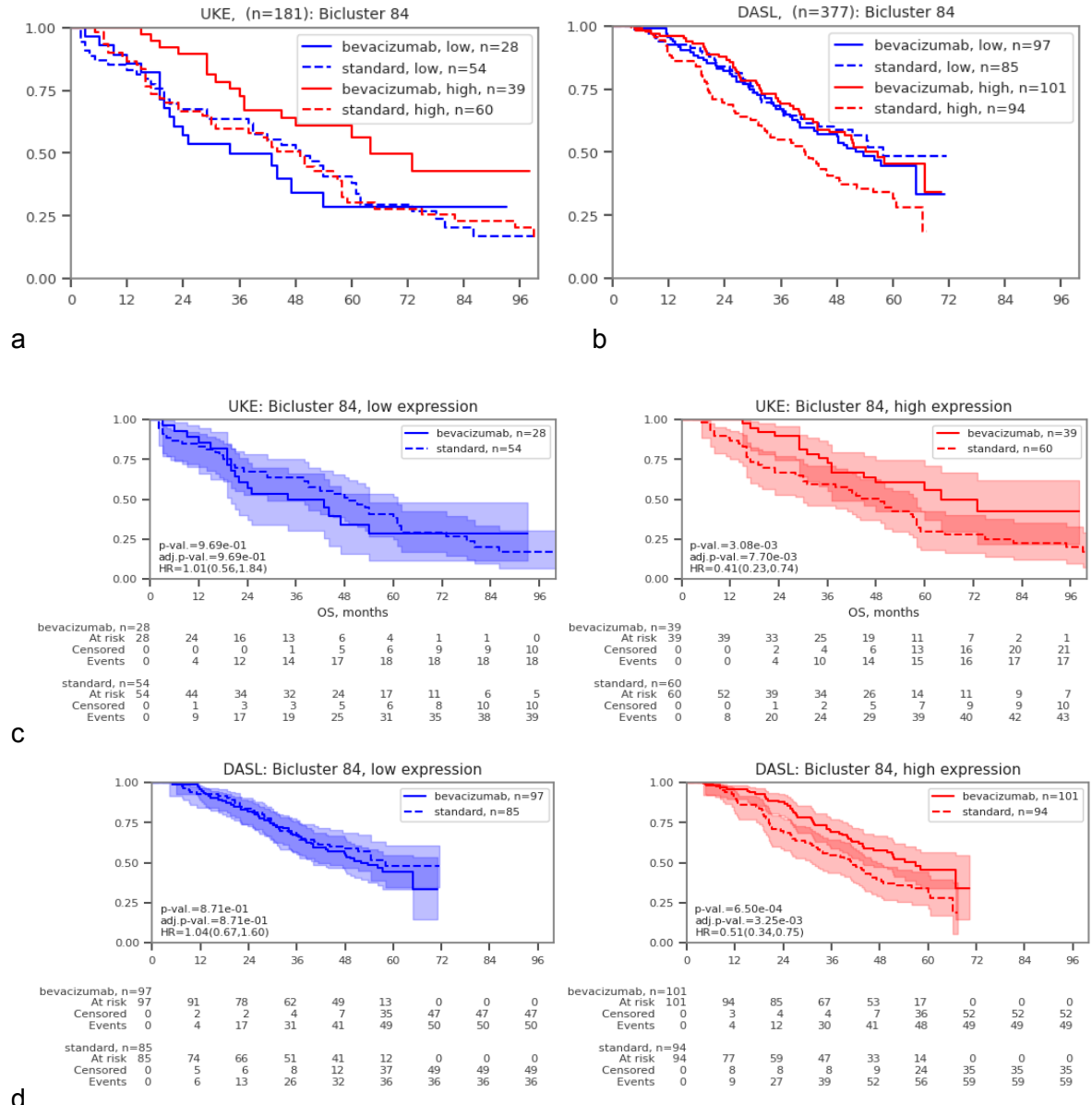

**Figure 3.** Kaplan-Meier plots for OS in the bevacizumab and standard treatment groups of the UKE (**a**) and DASL (**b**) cohorts stratified by expression of bicluster 84 genes. **c-d.** The comparison of OS under standard (dashed lines) and bevacizumab (solid lines) in the subgroups stratified by bicluster 84 expression. Hazard ratios (HR) estimate the risk ratios of death between the bevacizumab-treated group and the standard treatment group. In both cohorts, patients whose tumors over-expressed (red) bicluster 84 genes demonstrated longer OS when bevacizumab was added to standard chemotherapy. The OS of patients with tumors with opposite expression patterns was not significantly different between treatment groups. Since each of five bicluster candidates were tested independently, the resulting p-values were adjusted following the Benjamini-Hochberg procedure.

## Bicluster 84 signature replicates across datasets and platforms

Bicluster 84 consisted of five genes in the UKE cohort, of which two, the solute carrier organic anion transporter family member 6A1 (SLCO6A1) and the parathyroid hormone 2 receptor (PTH2R), were also measured in the DASL. Three other genes encoding a pseudogene (ENSG00000249721) and long non-coding RNAs (LINC00491, LINC01456), were absent from the DASL data.

Because UnPaSt is a probabilistic method which constructs biclusters from genes defining the sharpest sample clusters, many genes satisfying commonly used threshold of 0.05 for statistically significant differential expression between bicluster and background sample groups but having low SNR, may be not included into biclusters. Therefore, to better understand biological roles of bicluster 84 signature and to connect it with bevacizumab response, we extracted all genes statistically significantly differentially expressed between patient subgroups defined by this bicluster. To reduce the possible false findings, we focused on genes which were significantly differentially expressed in both UKE and DASL cohorts. Out of 307 and 49 genes significantly differentially expressed (|log2(FC)|>1, adj.p-value<0.05) between patient groups stratified by bicluster 84 in the UKE and DASL cohorts respectively, 14 genes were shared (Figure 4a,b;Supplementary Table S3). This overlap of two gene sets is stronger than expected by chance (one-sided Fisher's exact test p-value=1.79e-19, 15583 genes present in both cohorts were taken into account).

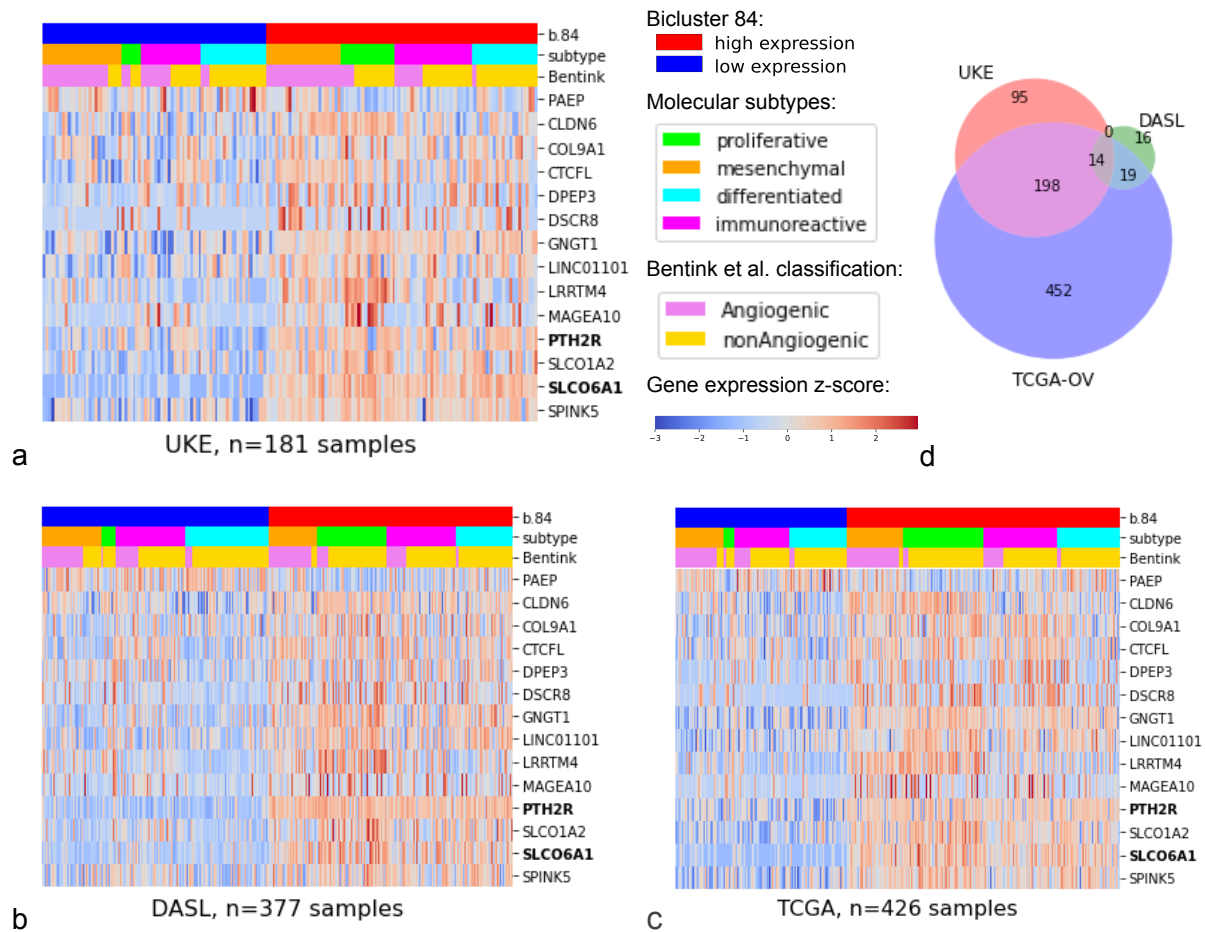

**Figure 4.** Normalized expressions of the 14 genes significantly differentially expressed by bicluster 84 signature (adj.p-value <0.05, |log2FC|>1) in the UKE (**a**), DASL (**b**), and TCGA-OV (**c**) cohorts. Gene included into bicluster 84 discovered in the UKE dataset are highlighted with bold text font. Red and blue bars on the top of the heatmaps highlight samples over- and under-expressing bicluster 84 genes. Sample classification with known molecular subtype (green - proliferative, orange - mesenchymal, cyan - differentiated, magenta - immunoreactive) is shown with the colorbar on the top of the heatmap. ENSEMBL gene ids were mapped to HGNC symbols. **d.** The overlaps of significantly differentially expressed gene sets for bicluster 84 in the UKE, DASL, and TCGA-OV datasets.

Additionally, we validated the presence of bicluster 84 signature in another publicly available RNA-seq based ovarian cancer cohort which included samples from patients not treated with bevacizumab, TCGA-OV (Figure 4c). The list of 683 genes differentially expressed in the TCGA-OV demonstrated even stronger overlap with the genes from the UKE dataset (212 shared genes; p-value=6.30e-269; 22491 genes presenting in both UKE and TCGA-OV were taken into account, Figure 4d). The better concordance between the expression signatures in UKE and TCGA-OV compared to DASL, may be attributed to the use of microarrays in the latter, whereas in the UKE and TCGA-OV gene expression was profiled by RNA-seq.

Interestingly, in all three cohorts, tumor subsets defined by bicluster 84 did not exclusively match any of four established molecular subtypes of EOC (Figures 4a-c). Tumours overexpressing this signature represented all four molecular subtypes in roughly equal proportions, although the majority of the samples classified as proliferative fall into the subgroup overexpressing bicluster 84 genes. These observations suggest that bicluster 84 may correspond to biological processes unrelated to molecular subtypes and define an additional classification of tumours. To better understand its biological role and find possible connection with the response to bevacizumab, we reviewed the fourteen genes included into this signature in all three datasets, and compared it with known pathways, functionally similar and tissue specific gene sets.

## Characterization of bicluster 84 signature

Among the fourteen genes significantly differentially expressed in tumor subgroups defined by bicluster 84, only PTH2R has been previously directly linked with angiogenesis. PTH2R is a selective parathyroid hormone receptor normally expressed at high levels in the brain and central nervous system and described as a key mediator of nociception and maternal behavior[26,27]. Three endogenous ligands have been described for PTH receptors: tuber-oinfundibular peptide of 39 residues (TIP39), PTH, and PTH-related peptide (PTHrP)[28]. PTH2R is also involved in the wound healing process, where ligand TIP39-mediated signaling induces the expression of decorin and the maturation of the extracellular matrix[29]. Also, PTH and PTHrP have been associated with angiogenesis activation[30,31]. There are few reports about the role of PTH2R in cancer, and interestingly among them a recent work identified a significant up-regulation of PTH2R in ovarian cancer cells and could further demonstrate its association with tumor cell proliferation, invasion and migration[32].

The SLCO6A1 gene encodes a protein called OATP6A1, which belongs to the solute carrier organic anion transporter family and has been identified as a cancer-testis antigen in lung tumor cell lines[33] and tumors of the lung, bladder, esophagus, and brain. Also LINC00491 has been implicated in various cancer types such as pancreatic cancer and lung cancer, promoting cell growth and metastasis[34,35].

Besides PTH2R and SLCO6A1, twelve other genes were differentially expressed in tumors with overexpression of bicluster 84 signature with an absolute log2-fold-change above 1 in both datasets (Supplementary Table S3). Same as SLCO6A1, three of them were cancer germline genes: DPEP3, MAGEA10, and CTCFL. The latest, also known as BORIS, encodes a master transcription factor that maintains cancer stemness by regulating several oncogenes and also other CTAs in different cancer types [36–40]. CTCFL is strongly expressed and frequently amplified in ovarian and other cancers[41], where it regulates VEGF-A expression alongside other factors[42]. Recently, it has been shown to maintain treatment-resistant phenotype of cancer cells[43]. Two extracellular matrix-related genes COL9A1 and CLDN6 were previously shown to be overexpressed in the C5 subtype of EOC defined by Tothil et al.[44]. Also, COL9A1, which encodes collagen IX, has been previously shown to exert an anti-angiogenic function, by increasing the levels of antiangiogenic TSP-1 and matrilin-1 and decreasing the levels of proangiogenic active MMP-9[45]. SPINK5 is a serine protease inhibitor, contributing to the anti-inflammatory and/or antimicrobial protection of mucous epithelia and described as a tumor suppressor[46]. Patients with renal cell carcinoma classified as responders to atezolizumab plus bevacizumab therapy[47]. GNGT1, LRRTM4, the long non-coding RNAs LINC01101 and DSCR8 have been previously discussed in the context of cancer, but to our knowledge, no association with angiogenesis has been reported so far.

The only gene demonstrating an opposite expression pattern to the other 13 genes of this signature, PAEP, encodes a 28-kDa glycoprotein glycodelin that participates in tumor angiogenesis by promoting endothelial cell proliferation, migration and tube formation through modulation of VEGF-A expression[48]. Glycodelin has been previously shown to be increased in plasma of patients with gynaecological malignancies, including ovarian cancer where it was correlated with poor prognosis at advanced stages[49]. Why the patients who respond to anti-angiogenic therapy have low levels of the angiogenic factor glycodelin PAEP remains to be clarified.

Although we validated the presence of this expression signature in three independently generated datasets, both over-representation and GSEA analyses provided limited insight into its biological role. Gene set over-representation analysis (ORA) of the 13 upregulated genes revealed small but significant overlaps with pathways related to "male testis" factors, confirming the aforementioned presence of cancer/testis antigens (CTCFL, DPEP3, MAGEA10), transmembrane transport activity (GO:0043252;GO:0015347;SLCO1A2 and SLCO6A1), anti-inflammatory response favoring leishmania infection (R-HSA-9662851; DPEP3, GNGT1, PTH2R), and external encapsulating structure organization (GO:0045229; COL9A1 and SPINK5) (Supplementary Table S4A). GSEA analysis confirmed the enrichment of genes specific to testis tissue among the genes differentially expressed in tumors stratified by bicluster 84 in the UKE dataset, but no associations passed the adjusted p-value cutoff of 0.05 for the DASL and TCGA-OV datasets (Supplementary Table S4B).

## Bicluster 84: the CTCFL signature

Since both ORA and GSEA analyses showed that known gene sets are not sufficient to explain bicluster 84 signature, we employed CORESH (see Methods) to search the Gene Expression Omnibus (GEO) database for public datasets where these genes are co-regulated. Among 44259 human-derived datasets, CORESH identified 128 where the bicluster 84 genes were co-regulated stronger than expected by chance (adj. p-value < 0.05; Supplementary Table S5A). Text mining analysis of the titles and descriptions revealed the overrepresentation of the word "pluripotency" (adj.p-value=3.8e-24) and other terms associated with stemness among the top-ranked datasets (Supplementary Table S5B;Methods), pointing to a potential key role of CTCFL/BORIS behind this signature. The acquisition of stemness properties by cancer cells promotes their growth, invasiveness, and increases their drug resistance. A link between cancer cell stemness and angiogenesis activation, e.g. through secretion of angiogenic factors such as VEGF-A, has been described for CSCs in general and for ovarian cancer stem cells in particular[50–52]. Therefore, an overexpression of this signature in the tumor at the beginning of treatment may be associated with the effectiveness of anti-angiogenic therapy. Furthermore, Krishnapriya et al. described the contribution of cancer stem cells in high-grade serous ovarian cancer to the formation of blood and lymphatic vessels. Spheroids derived from ascites-derived cells cultured in stem cell medium showed high expression of endothelial, pericyte and lymphatic endothelial markers and were able to form endothelial and lymphatic tube-like structures *in vitro* and *in vivo*. Notably, bevacizumab inhibited the differentiation of spheroids into endothelial cells *in vitro*. These results suggest that cancer stem cells may themselves contribute to angiogenesis and lymphangiogenesis in serous ovarian cancer and that this process can be inhibited by anti-angiogenic treatment with bevacizumab[53].

## Other biclusters potentially predictive for OS

Biclusters 70, 90, 109 and 130 were selected by the random forest algorithm but were not associated with statistically significant differences in OS. However, because these patterns were replicated in three independent EOC cohorts (Figure S7,S8), and patient subsets stratified by them demonstrated similar trends in OS in both UKE and DASL cohorts (Figure S5), they have potential relevance for further investigation and re-evaluation using larger patient cohorts.

Interestingly, in all three cohorts, stratification of samples by bicluster 70 is strongly correlated with the pro- and anti-angiogenic classification of EOC proposed by Bentink et al., 2012 [54] (Figure S8). Consistent with their findings, tumors overexpressing these genes are predominantly classified as mesenchymal and immunoreactive in all three cohorts. Bentink et al. also hypothesized that the presence of this angiogenic signature could inform decisions on bevacizumab administration. Paradoxically, our results suggest that, in both UKE and DASL cohorts, patients whose tumors overexpress the pro-angiogenic signature benefit less from the addition of bevacizumab in OS (Figure S5e,f) and shows decreased PFS compared to low-expression group (Figure S6a,b). This finding suggests that these tumors may not only rely on VEGF to activate angiogenesis, but instead exploit alternative angiogenic pathways thus not being sensitive to bevacizumab. It is well known that there is a plethora of growth factors and pathways that tumor cells can exploit to induce angiogenesis and in several cancers these alternative angiogenic pathways have been described to be associated with reduced tumor susceptibility to anti-VEGF agents[55].

The bicluster 130 signature (Figures S8a-c) is enriched with genes specific to the adipose tissue and related to lipid metabolism (e.g. ADIPOQ, CIDEA, CIDEC, FABP4). In both cohorts, bevacizumab-treated patients whose tumors overexpressed this signature demonstrated decreased OS (Figure S5c,d) and PFS (Figure S6k,l) compared to those who also received bevacizumab but were negative for this signature. Also, most of the tumors overexpressing this signature are classified as mesenchymal and from about one third to more than a half of all mesenchymal tumors overexpress these genes. It is not clear why increased lipid metabolism, and mesenchymal phenotype might correlate with a poorer response to anti-VEGF treatment. Since the set of genes significantly differentially expressed genes in the UKE and DASL datasets is also enriched with lymphoid cell infiltration markers (e.g., CCR7, CD79B, CST7, EOMES, FCRLA, IL1RL1, ITK, TNFRSF13B), one could speculate that in these tumors, angiogenesis is actively controlled by immune cells which secrete a high variety of angiogenic factors (e.g. EGF, FGF2, CXCL8, CXCL12, TNFα and semaphorin 4D12[56]) which help a tumor to overcome VEGF blockade more easily. Another gene strongly over-expressed in these tumors, CD36, represents a known negative regulator of angiogenesis and its over-expression in the tumor prior to treatment may reflect the pre-existing inhibition of the VEGF signal[57,58] making the addition of bevacizumab inefficient.

## Impact of platform on reproducibility of expression signatures

While evaluating the biclusters in the TCGA-OV data, we noticed that in general, biclusters discovered in the UKE dataset replicated in this RNA-seq-based cohort much better than in the DASL (Figure S9). All 23 biclusters replicated in the DASL data were also found in the TCGA-OV, and many biclusters that failed to replicate in the DASL were successfully identified in the TCGA-OV. Additionally, biclusters replicated in the TCGA-OV retained more genes than their counterparts in the DASL. This higher concordance between UKE and TCGA-OV is likely due to both cohorts being profiled using RNA-seq technology, whereas DASL is microarray-based and lacks or imprecisely measures some genes. One of the most illustrative examples of this limitation is bicluster 129, consisting of 128 genes in the UKE data, the majority of which encode immunoglobulin or T-cell receptor chains. In the DASL cohort, only 19 out of 129 genes were measured and only 6 were included into the bicluster. In contrast, the majority of the 129 genes presented in the TCGA-OV, and 80 were included into this bicluster. Interestingly, in all three cohorts, only about a half of samples included into this bicluster were classified as immunoreactive (Figure S10). Overall, 97 biclusters discovered in the UKE data and replicated in the TCGA-OV did not match any of four molecular subtypes (adjusted Rand index < 0.25) were identified. Such reproducible expression patterns may have biological or clinical significance and might be of interest for future research. However, because the patients in the TCGA-OV cohort only received chemotherapy[23] we cannot draw any conclusion about the association of these with survival under bevacizumab treatment.

# Conclusion

Cytoreductive surgery and platinum/paclitaxel chemotherapy followed by maintenance therapy with PARP inhibitors (PARPi) and/or the anti-angiogenic drug bevacizumab are the standard treatment for advanced ovarian cancer. However, no reliable biomarkers to guide the decision on maintenance therapy have yet been proposed to date.

To address this urgent need for bevacizumab response biomarkers, we explored a novel dataset including expression profiles and clinical data of EOC patients treated at Universitätsklinikum Hamburg-Eppendorf (UKE). Although our analysis of the UKE dataset did not confirm associations between molecular subtypes of EOC and bevacizumab response in terms of OS or PFS, biclustering analysis revealed multiple expression signatures which replicated in independently generated EOC expression datasets (DASL, TCGA-OV) and were unrelated to known molecular subtypes. One of them, bicluster 84, was prioritized by random survival forest as a predictive biomarker candidate associated with OS under bevacizumab treatment. Although its interaction with treatment in Cox regression model adjusted for patient age, tumor stage and surgery outcome, did not pass the commonly used threshold for statistical significance (UKE:p-val.=6.92e-02, adj. p-val.=0.23; DASL:p-val.=1.36e-02, adj.p-val.=0.07), probably due to lack of power, stratified analysis suggests its potential prognostic value. For patients whose tumors overexpressed this signature, bevacizumab addition to standard treatment significantly prolonged OS in both UKE (HR=0.41(0.23-0.74), p-value=3.08e-03; adj.p-value=7.70e-03) and DASL (HR=0.51(0.34-0.75), p-value =6.50e-04; adj.p-value=3.25e-03) cohorts. In the rest of patients, no significant differences in OS explained by bevacizumab addition were observed. Further, because traditional functional over-representation approaches relying on known gene sets aided limited support in finding biological interpretation of this expression signature, we performed mining of public expression data from NCBI GEO and identified connection of this signature with the acquisition of stemness properties. We hypothesize that this expression signature is driven by CTCFL/BORIS, which is frequently amplified in ovarian and other cancers[41], and has been shown to regulate VEGF-A expression[42] and to maintain stemness and treatment-resistant phenotype of cancer cells[43]. In summary, this signature not only complements the existing molecular classification of EOC, and represents a promising biomarker candidate for bevacizumab response, but may also inform the choice of therapy targeted against CTCFL[59,60].

It is important to note that our models only partially explained the observed differences in survival between treatment groups. Our analysis indicates that in both the discovery and validation cohorts, different molecular and clinical variables are more relevant for prediction of OS than PFS. OS is considered the gold standard endpoint in cancer clinical trials because it can be accurately measured from the documented dates of death, and has a low risk of reporting bias[61]. In contrast, PFS, though also an objective primary endpoint, may be more susceptible to observational bias. Particularly in ovarian cancer, the limitations of imaging systems, but also of the main biomarker CA125, and the different timing of assessment contribute to the complexity of using PFS as a surrogate endpoint for OS in ovarian cancer and highlight the need for careful consideration when interpreting results based on PFS alone.

Furthermore, given the high molecular heterogeneity of EOC and its unexplored diversity beyond the four molecular subtypes, it is possible that bevacizumab efficacy may be determined by multiple molecular features of a tumor with little individual contributions. Therefore, to comprehend this heterogeneity and to develop a stronger predictor of bevacizumab response, replication of this study with larger cohorts is required.

Besides the limited cohort size, another challenge in our analysis was the difficulty of reproducing expression signatures discovered in RNA-seq data in the validation dataset generated using microarray technology. The absence of certain genes in the validation cohort complicated the evaluation of bicluster candidates by disabling robust sample clustering. This highlights the need for an independent validation cohort of EOC patients treated with bevacizumab whose tumor expressions would be profiled using RNA-seq technology.

# Methods

## UKE and DASL cohorts

The UKE cohort was established at the Department of Gynecology, University Medical Center Hamburg Eppendorf (UKE), Hamburg, Germany. Cryopreserved tumor tissue samples (n=244) corresponding to 212 patients with peritoneal, fallopian tube and ovarian carcinoma treated at this center between 2009 and 2017 were available in the gynaecological biobank and were included in this cohort. All patients gave written informed consent for biomaterial to be collected and for access to their clinical records. The procedure was approved in advance by the Ethics Committee (Ethik-Kommission der Ärztekammer Hamburg, No. 190504). Patient data were followed from the date of initial diagnosis until 2022. Detailed information on the RNA isolation procedure and RNAseq analysis is provided below. The characteristics of the entire reduced UKE cohort used in the current study are shown in Table 1.

The DASL cohort has been described previously[17]. Briefly, paraffin-embedded tissue was available from 423 patients with primary ovarian, fallopian tube or primary peritoneal cancer from the AGO-OVAR11 trial who were initially included in this cohort. For further expression analysis, 380 samples with both expression and clinical data available were included.

## RNA isolation from ovarian cancer tissue and RNAseq analysis

RNA isolation from 244 tumor tissue samples corresponding to 212 ovarian cancer patients was performed as previously described[62,63]. Briefly, tissue samples obtained during cytoreductive surgery were immediately stored in liquid nitrogen as fresh frozen samples. Haematoxylin-eosin-stained cryo-sections were performed to evaluate the histological characteristics and tumor content of each sample. Subsequently, 20–30 cryosections (approximately 16 µm) were disintegrated using the Precellys homogeniser (WVR International GmbH, Darmstadt, Germany) and RNA was extracted using the RNAeasy Kit (Qiagen GmbH, Hilden, Germany), according to the manufacturer's instructions. RNA quantity and integrity were assessed using a Bioanalyzer device (Agilent, Santa Clara, CA, USA). RNA sequencing was performed by BGI Genomics (Shenzhen, China) using the DNBseqTM Technology Platform. Sequence reads were processed with fastp (v0.23.2) to remove sequences originating from sequencing adapters and sequences of low quality (Phred quality score below 15)[64]. Reads were then aligned to the human reference assembly (GRCh38.106) using STAR (v2.7.10a)[65]. Gene-level raw read counts were normalized using the median ratio method [66] implemented in DESeq2 (v1.34.0)[67] R package and log2(x+1) transformed. Genes with 15 or more normalized counts in less than 10 samples were considered weakly or infrequently expressed and were excluded from further analysis.

## Expression data

Transcriptome data and corresponding clinical and histopathological information from three independently created ovarian cancer patient cohorts were analyzed in this study. For two of them, data was publicly available: DASL (GSE140082)[17] and TCGA-OV[23].

Normalized and log-transformed gene-level expressions from the DASL cohort comprising 380 samples preprocessed as described in Kommos et al. 2017[17] were provided by Stefan Kommoss. Three samples were excluded from the analysis because information on tumor stage or operation result was missing.

RSEM-normalized and log2-transformed gene-level read counts and patient data including OS and PFS time and events, tumor stages for TCGA-OV cohort[23] were obtained from XENA[68] TCGA Pan-Cancer (PANCAN) data hub (https://xenabrowser.net/datapages/?cohort=TCGA%20Pan-Cancer%20(PANCAN)).

The UKE dataset comprises 244 tumor samples obtained from 212 individuals diagnosed with EOC. Gene-level read count data is publicly available at NCBI GEO[69] (link) under accession XXXXXX. Normalized expression profiles of all 244 samples were used in unsupervised analysis to enable more robust sample clustering and facilitate the detection of infrequent sample subsets and, but only 181 samples were taken into account in survival analysis. When multiple samples received from one patient were available, only one non-metastatic sample with the highest tumor purity was kept in survival analysis (n=28 samples were excluded from survival analysis based on this criteria). Also, samples obtained from patients who received neoadjuvant or treatment different from standard chemotherapy or bevacizumab (n=17), had recurrent EOC (n=7), or for whom information on survival, therapy, age at diagnosis, FIGO and/or residual tumor after surgery was not available (n=11) were excluded from survival analysis.

## Identification of known molecular subtypes

Molecular subtypes were assigned using a random forest-based classifier implemented in the *consensusOV*[19] R package v1.24.0, *get.subtypes()* function was applied to normalized, log2(x+1)-transformed and standardized expression data with default parameters. Consensus molecular subtypes predicted for the DASL and TCGA-OV cohorts were similar to previously published classifications (Supplementary Table S1B-C). Additionally, to ensure that the predicted molecular subtypes consistently express biomarkers described in the literature, based on the lists used by Verhaak et al.[21] and Talhouk et al.[24] we compiled a list of 98 subtype-specific biomarkers (18 for proliferative, 27 for mesenchymal, 16 for differentiated, and 37 for immunoreactive; Supplementary Table S1D) and verified their expression in three cohorts UKE, DASL and TCGA-OV (Supplementary Figure S11).

## Biclustering for unsupervised patient stratification

Unsupervised patient stratification was performed by UnPaSt[25] (https://github.com/ozolotareva/unpast) v0.1.11, and applied to between-sample normalized and log2-transformed expression data with default parameters. UnPaSt standardizes each gene in the expression data matrix and searches for differentially expressed biclusters – submatrices consisting of genes concordantly over- or under-expressed in a subset of samples. Each such bicluster stratifies all samples into two subpopulations differentially expressing a certain gene set (Fig. 5a). In other words, UnPaSt identifies all gene sets in transcriptome data which define two well-separable sample groups, and cluster samples independently within each such gene set. Therefore, in contrast to conventional sample clustering (Fig. 5b), UnPaSt reveals overlapping sample clusters and simultaneously extracts their specific genes, thus providing a more comprehensive and interpretable picture of data heterogeneity (Fig. 5c).

Since UnPaSt is a probabilistic method, consensus biclusters constructed based on the results of five independent runs were analyzed. For building consensus biclusters, the same approach as described in Hartung et al., 2024[25] was used. To enable the detection of biclusters representing rare patients subgroups, the complete UKE dataset comprising 244 tumor samples was input to UnPaSt run with default parameters. Out of 234 biclusters, 177 biclusters defining two well-separable sample clusters (average signal-to-noise ratio across all bicluster genes (SRN) > 1.5), each including at least 10 samples from both treatment groups were selected for evaluation as predictive and prognostic biomarkers. After the detection, only 181 samples selected for survival analysis were taken into account.

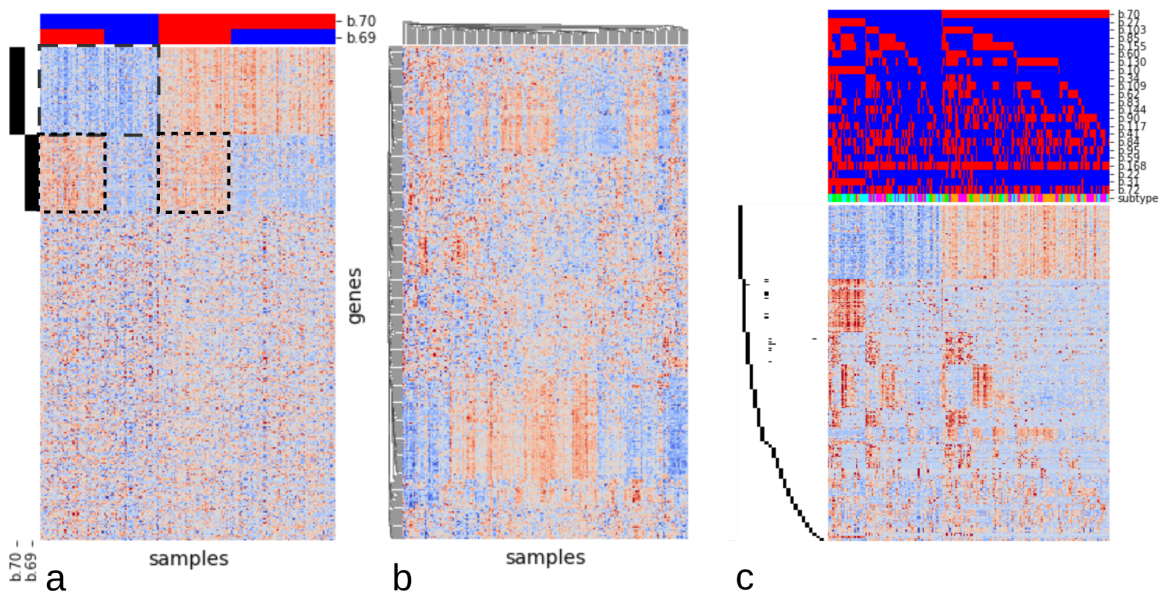

**Figure 5**. An example of biclustering (**a**) and clustering (**b**) analysis of 327 genes from the UKE dataset. For illustration purposes, 109 genes from two biclusters identified by UnPaSt and 219 randomly chosen genes were selected. **a**. Two biclusters identified by UnPaSt are highlighted with dashed frames. Gene membership in biclusters is shown with black bars on the left margin of the heatmap. Colorbars on the top of the heatmap show whether the bicluster genes are up- or down-regulated in each sample. Although a bicluster always divides samples into two subpopulations, we refer to a smaller subpopulation as "bicluster" and to a larger one as "background". If a bicluster divides all patients into two equal-sized subsets, the subset with higher expression is assigned to the "bicluster" group. Based on the direction of the expression pattern, up-regulated (e.g. bicluster 69), down-regulated (e.g. bicluster 70) biclusters can be distinguished. **b**. The results of hierarchical clustering of samples and genes with average linkage and Euclidean distance metrics. **c**. 23 biclusters identified in the UKE datasets which were replicated in DASL. Expression of 264 genes in 181 samples from the UKE dataset is shown, biclusters are ordered by the number of genes decreasing.

## Replication of biclusters discovered in the UKE data

To identify a bicluster in a new dataset that resembles a previously discovered bicluster, samples from the new dataset were divided into two clusters by k-means performed only for available bicluster genes in the new dataset. Individual genes from the new bicluster which poorly separated two sample clusters (SNR<0.5) were removed. Only new biclusters with at least 2 genes and 10 samples were kept.

## Random survival forest

Random survival forest models[70] from scikit-survival[71] version 0.23.0 Python package were used to predict OS and PFS in treatment subgroups of the UKE cohort. Model inputs included patients' age, tumor stage, surgical outcome, and expression levels of 23 biclusters, represented as binary variables.

For each treatment group from the UKE cohort, a random survival forest model was trained through a grid search approach with 5-fold cross-validation. Validation was performed within the corresponding treatment group of the DASL dataset. Individual feature importances were determined by averaging the decrease in model performance across 1000 permutations of the validation data, where the feature values were randomly shuffled[72].

## Statistical analysis

Time-to-event analysis and Kaplan-Meier curves were performed using *lifelines* Python package v0.27.0[73]. Prognostic and predictive power of each biomarker candidate was evaluated separately by fitting Cox Proportional Hazards models adjusted by donor's age, operation result, and tumor FIGO:

Survival ~ **treatment*biomarker** + biomarker + treatment + age + stageIIIC +stageIV + residual_tumor<1cm + residual_tumor>1cm.

Stage and operation results were defined as binary variables, distinguishing patients with residual tumor < 1cm, residual tumor > 1cm, and patients with FIGO stages IIIC and IV. Patients with macroscopically invisible residual tumors, and with tumor stages earlier than IIIC were considered as baseline risk groups. Molecular subtypes and biclusters were modelled with binary variables. Additionally, to evaluate the effect of bevacizumab and standard treatment in two subpopulations stratified by a molecular subtype, or a bicluster, the following regression models were fitted:

Survival~ **treatment** + age + stageIIIC +stageIV + residual_tumor<1cm +residual_tumor>1cm.

Differential expression analysis was performed using *limma v*3.58.1 for the microarray-based DASL dataset and using *limma-voom* for the RNA-seq data[74]. Genes with at least 10 normalized counts in less than 10 samples were removed from the UKE and TCGA-OV datasets. Genes with adjusted p-values < 0.05 were considered to be differentially expressed.

For gene sets for overrepresentation and GSEA analyses, *gseapy*[75] python package v1.1.3 and gene set libraries *GO_Molecular_Function_2023*, *GO_Biological_Process_2023*[76,77], *Reactome_2022*[78], *KEGG_2021_Human*[79,80], *WikiPathway_2023_Human*[81], *GTEx_Tissues_V8_2023*[82], *Descartes_Cell_Types_and_Tissue_2021*[83], *CellMarker_2024*[84], and *Tabula_Sapiens*[85] provided by *EnrichR*[86] were used. The subset of 20376 genes expressed in the UKE cohort and successfully mapped to HGNC gene names was used as the background. Up- and down-regulated gene sets were tested for over-representation separately. Only overlaps with gene sets comprising 5 to 200 genes were tested, and only overlaps of at least two genes passing adjusted p-value cutoff of 0.05 (Right-tailed Fisher's exact test) were taken into account.

To adjust p-values for multiple testing, Benjamini-Hochberg procedure[87] was performed.

## Signature-based public dataset search

To find datasets where genes from bicluster 84 were co-regulated we employed an unbiased dataset search approach described previously[88]. First, we gathered a compendium of 44259 public human gene expression matrices from NCBI GEO database[89], including microarray datasets available directly from NCBI GEO and RNA-seq gene count matrices obtained from ARCHS4[90] and DEE2[91] databases. For each of these datasets we ran Gene Set Co-regulation Analysis (GESECA) algorithm from the *fgsea* R package[92] using the 13 genes over-expressed in bicluster 84 as the input. The datasets then were sorted by GESECA p-value, so that the top hits would correspond to the lowest calculated p-values. Further, using the NLTK library we extracted terms associated with each dataset in the compendium.

Hypergeometric test was then used to identify words overrepresented within the top 300 hits. The described analysis pipeline is available as a web-service: https://alserglab.wustl.edu/coresh/.

## Code availability

https://github.com/ozolotareva/bevacizumab_ovca_signature

# Supplementary Figures

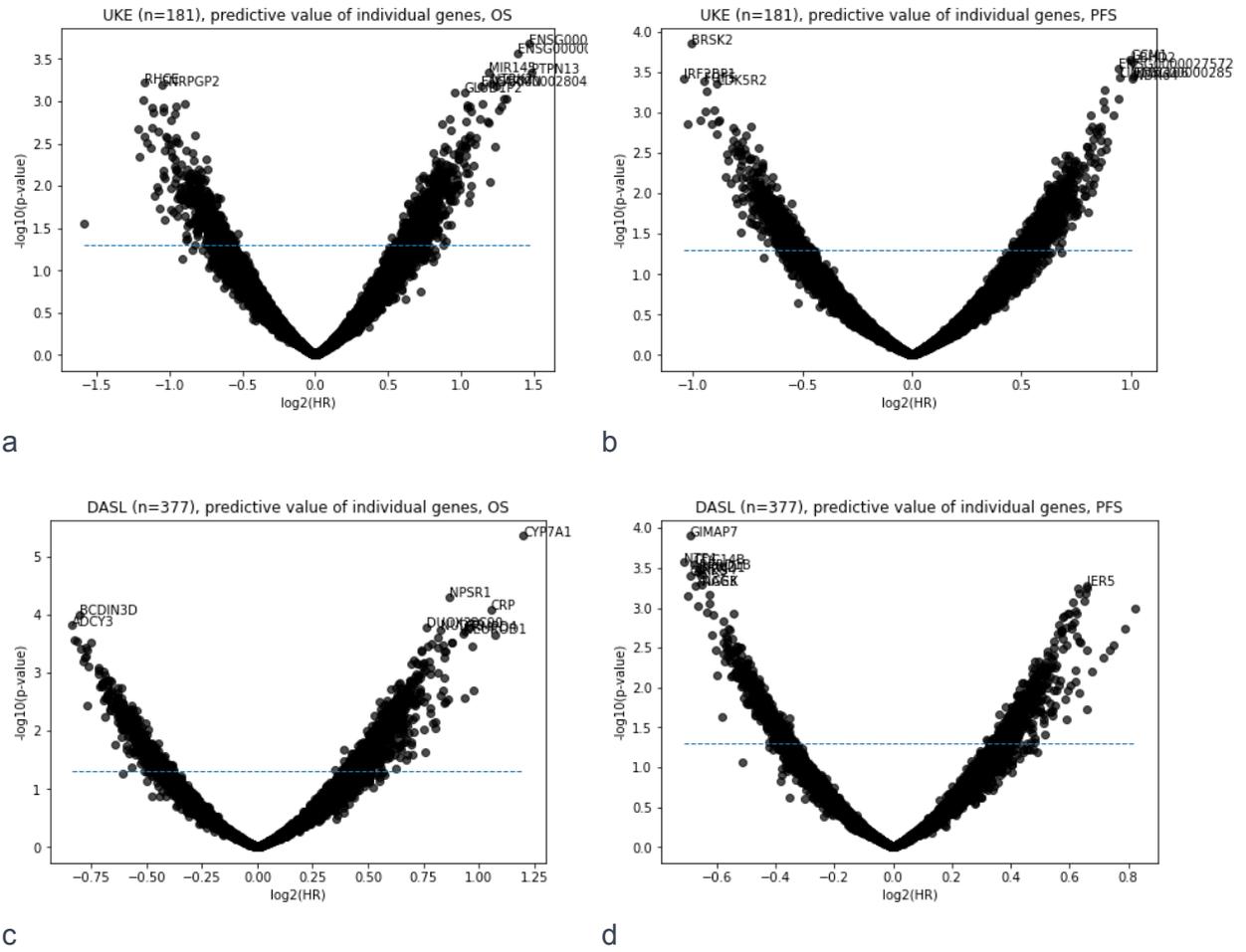

a

b

c

d

**Supplementary Figure S1.** Predictive values of individual gene expressions for overall survival (OS; **a,c**) and recurrence-free survival (RFS; **b,d**) in the UKE (**a,b**) and DASL (**c,d**) cohorts. Not adjusted p-values for interaction between treatment and gene expression are reported. After correction for multiple testing, all adjusted p-values exceeded 0.93 for both OS and PFS in the UKE, and 0.09 in the DASL.

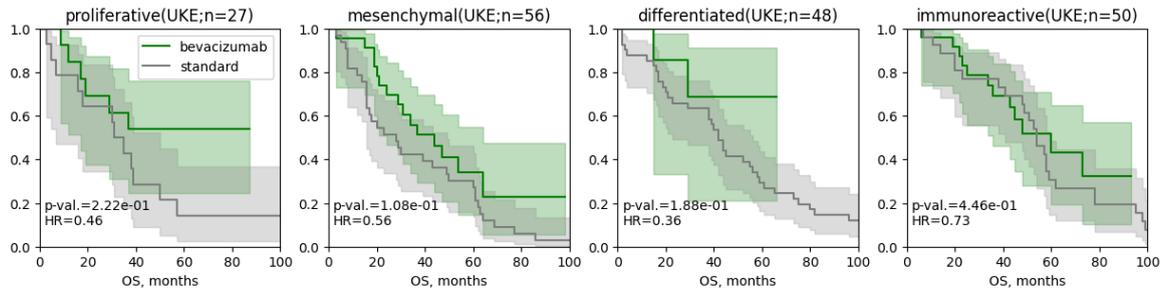

a

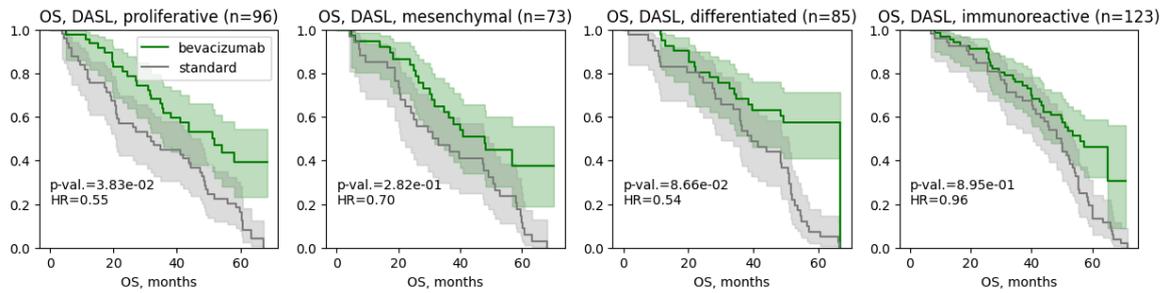

b

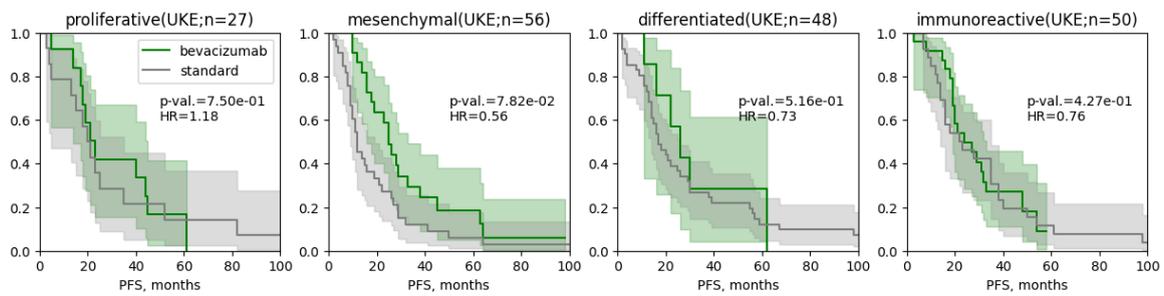

c

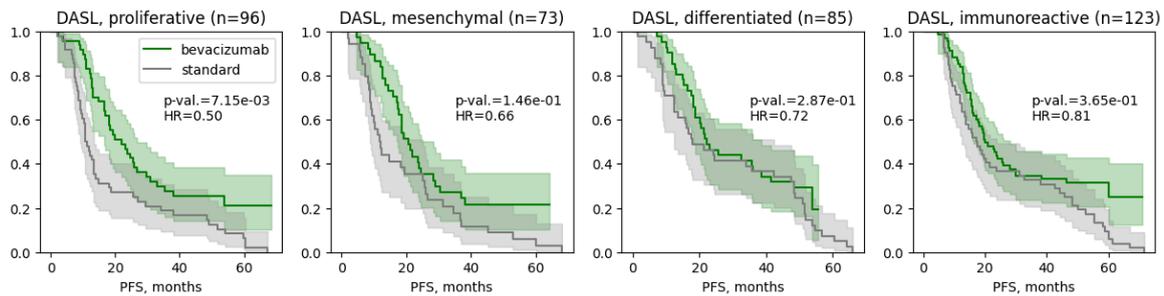

d

**Supplementary Figure S2.** Kaplan-Meier plots for OS (**A,B**) and PFS (**C,D**) in the bevacizumab and standard treatment groups in the UKE (**A,C**) and DASL (**B,D**) cohorts stratified by molecular subtype. In the UKE cohort, molecular subtypes were determined by the *consensusOV* classifier. In the DASL cohort, molecular subtypes reported by Kommos et al. were used. P-values not adjusted for multiple testing are reported.

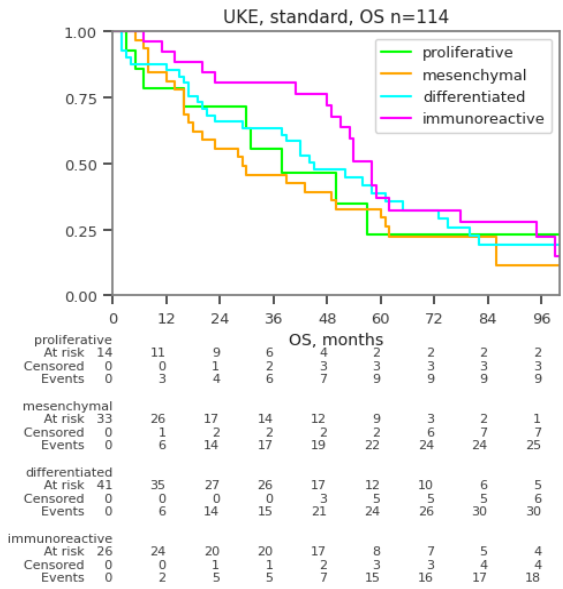
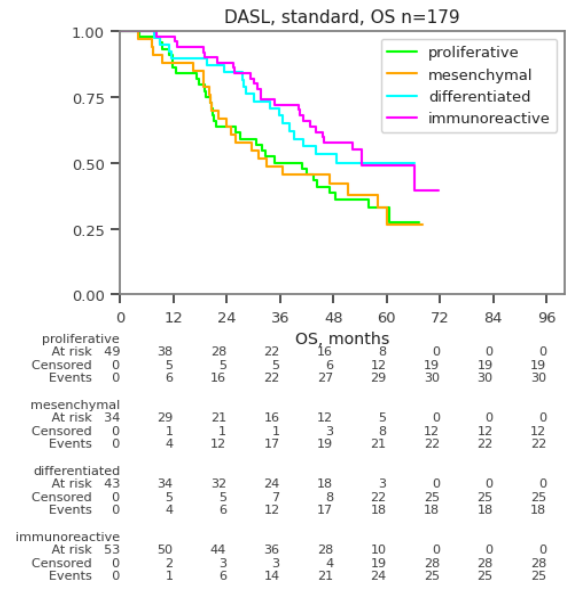

a
b

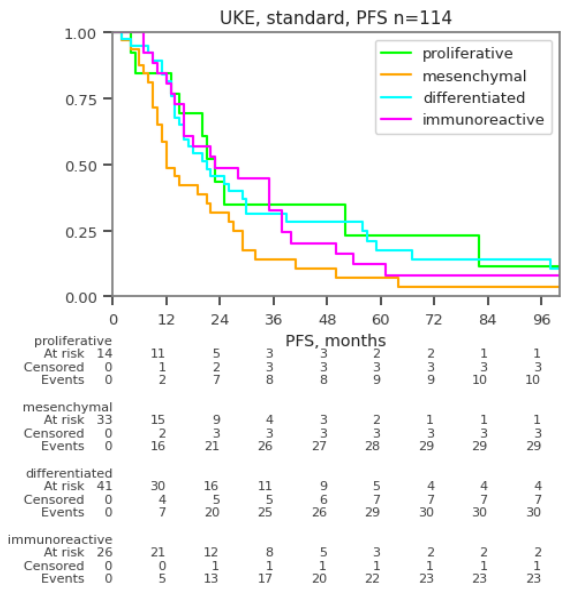
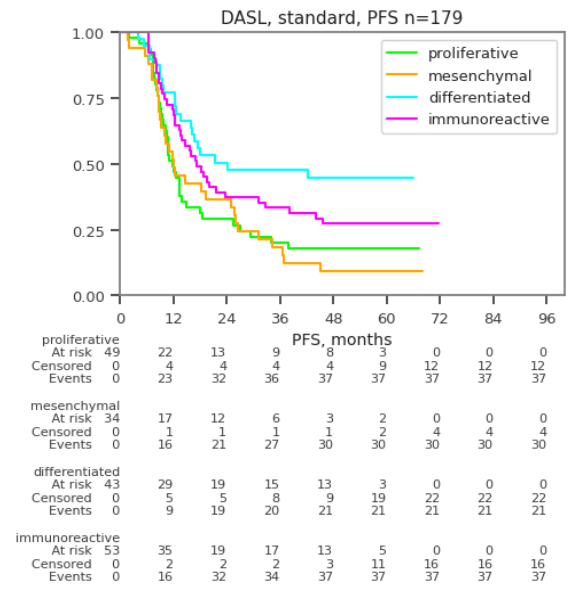

c
d

**Supplementary Figure S3.** Kaplan-Meier plots for OS (**a,b**) and PFS (**c,d**) for patients stratified by molecular subtype under standard treatment in the UKE (**a,c**), DASL (**b,d**).

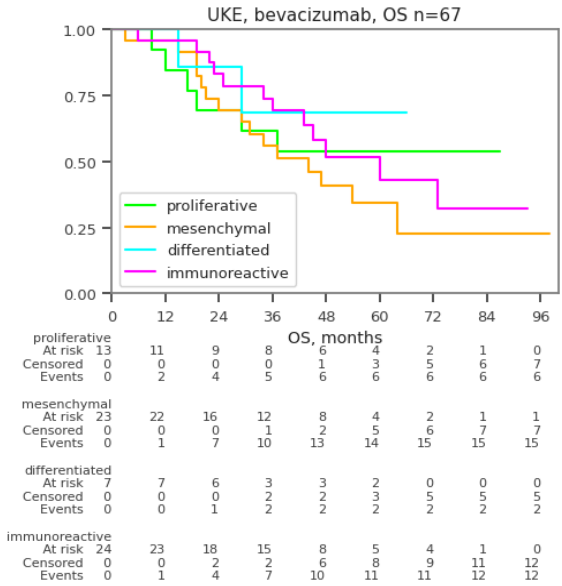
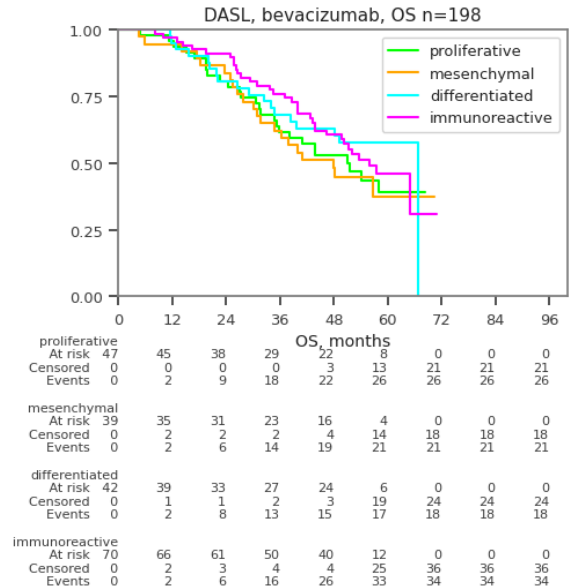

a

b

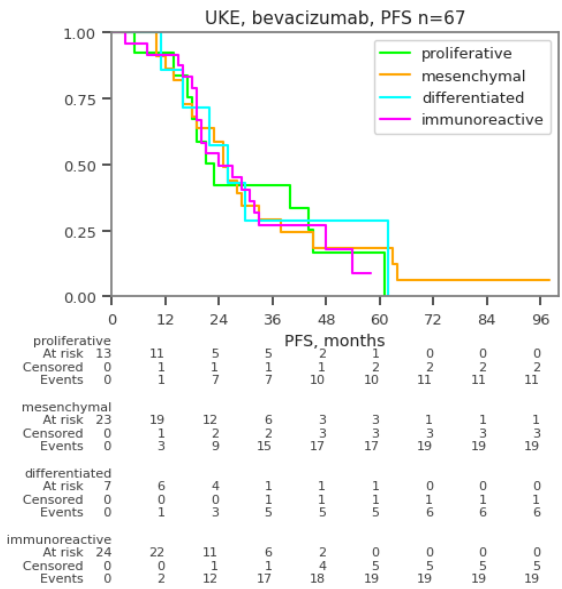
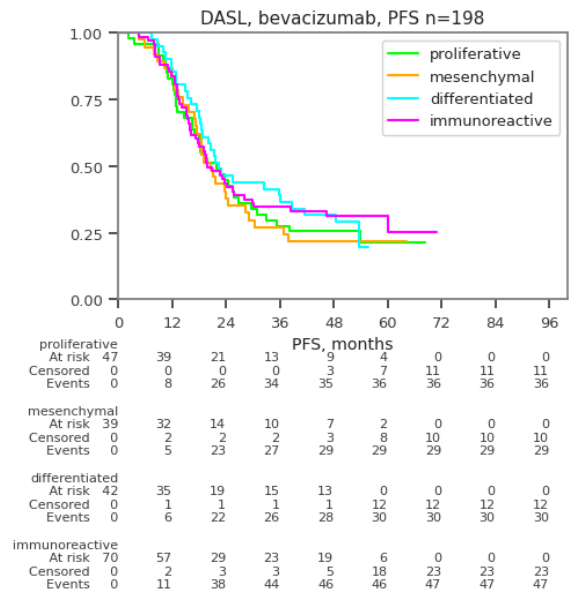

c

d

**Supplementary Figure S4.** Kaplan Meier plots for OS (**a**,**b**) and PFS (**c**,**d**) of patients stratified by molecular subtype under bevacizumab treatment in the UKE (**a**,**c**) and DASL (**b**,**d**) cohorts.

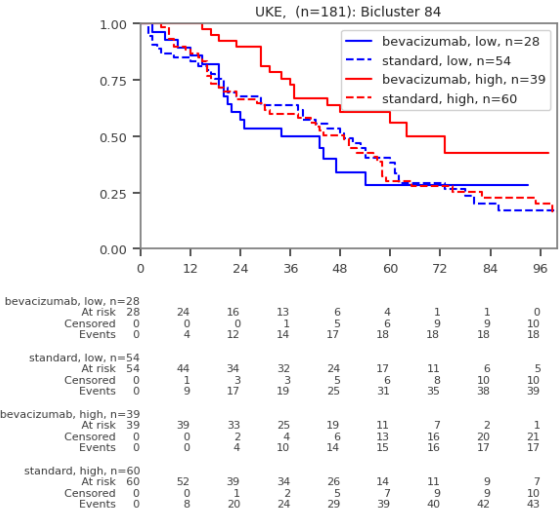

a

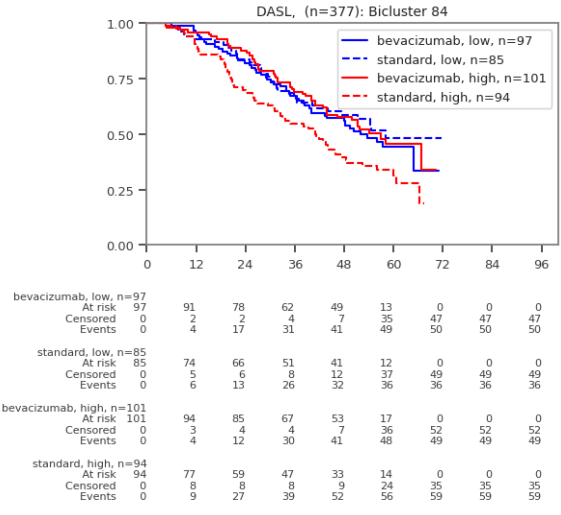

b

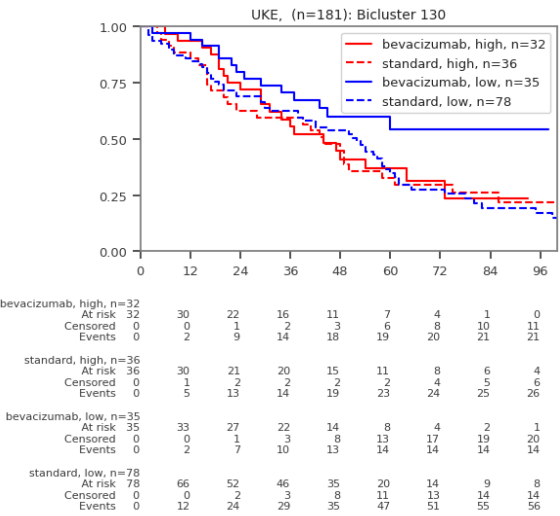

c

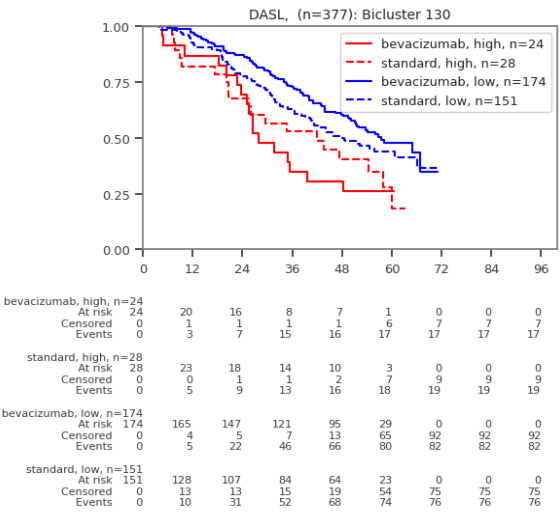

d

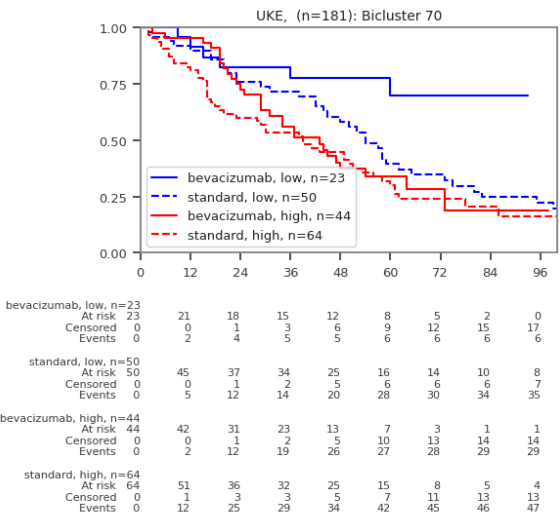

e

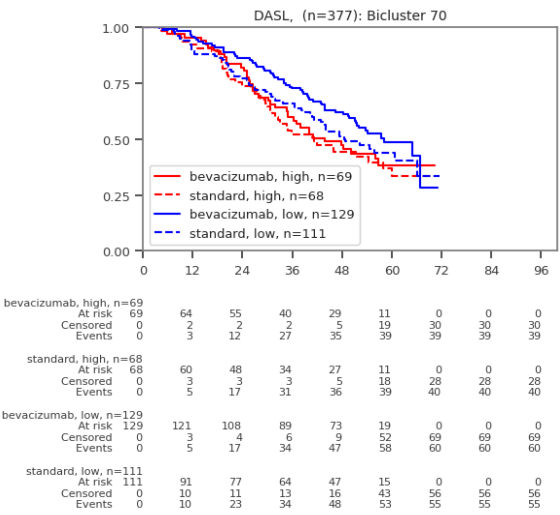

f

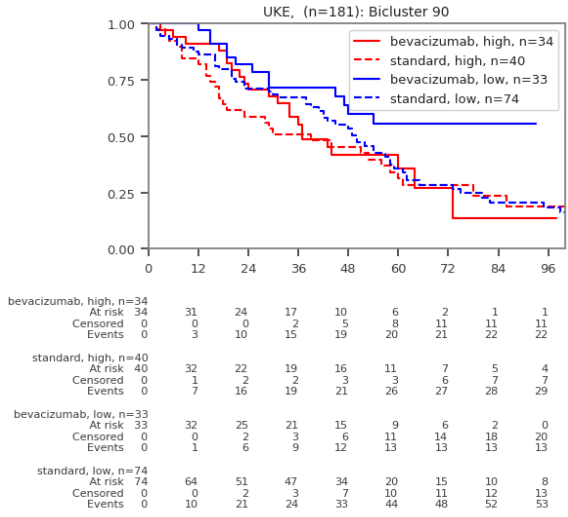

g

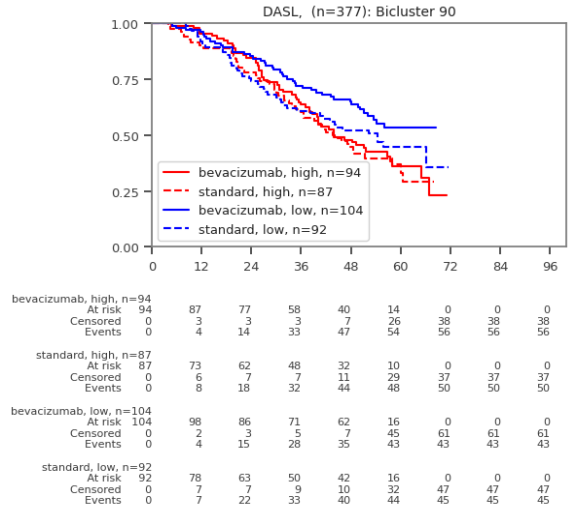

h

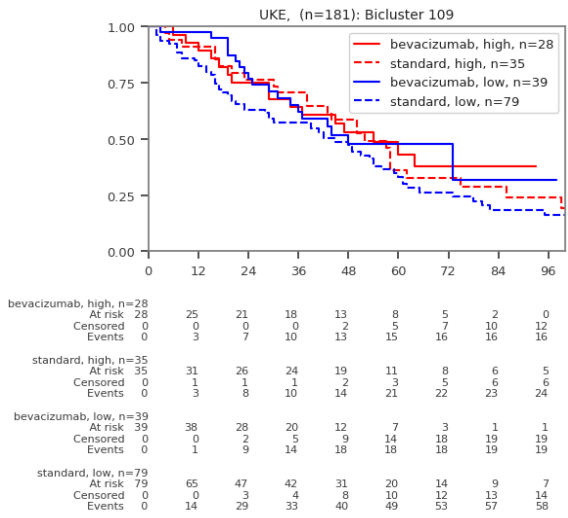

k

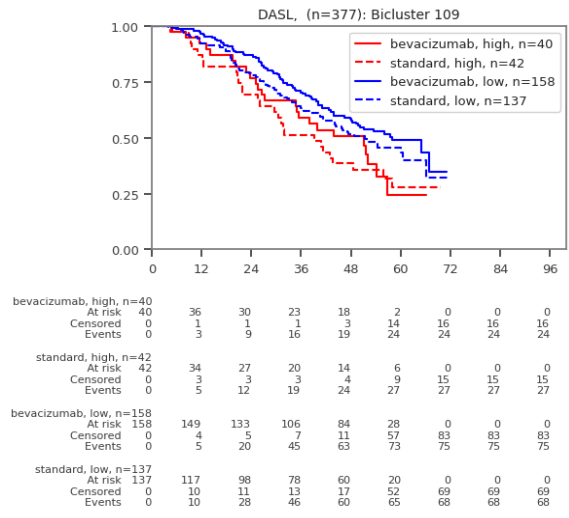

l

**Supplementary Figure S5.** Kaplan-Meier plots for OS of patient subgroups of the UKE and DASL cohorts stratified by treatment (solid line - bevacizumab; dashed line - standard chemotherapy) and expression of genes comprising biclusters 84 (**a,b**), 130 (**c,d**), 70 (**e,f**), 90 (**g,h**), and 109 (**k,l**) prioritized by random survival forests as the best predictors of OS.

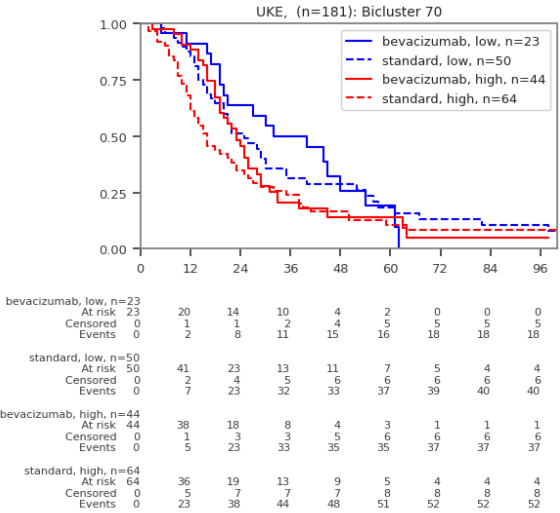

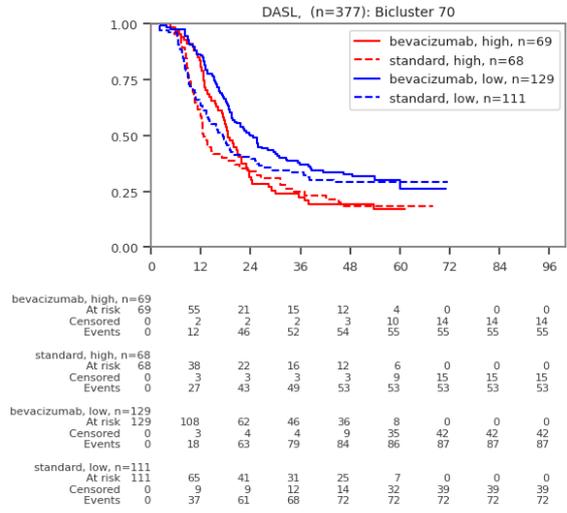

a

b

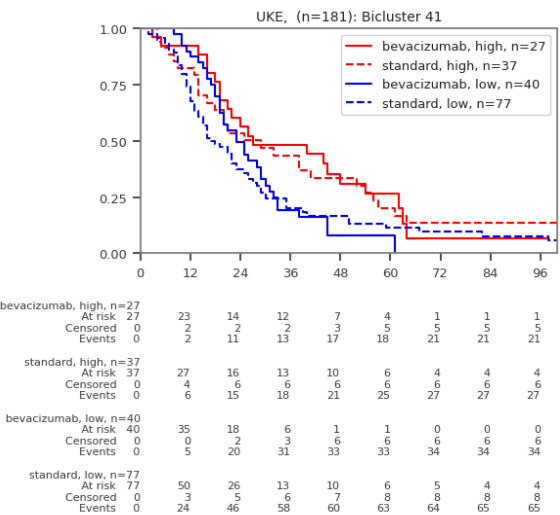

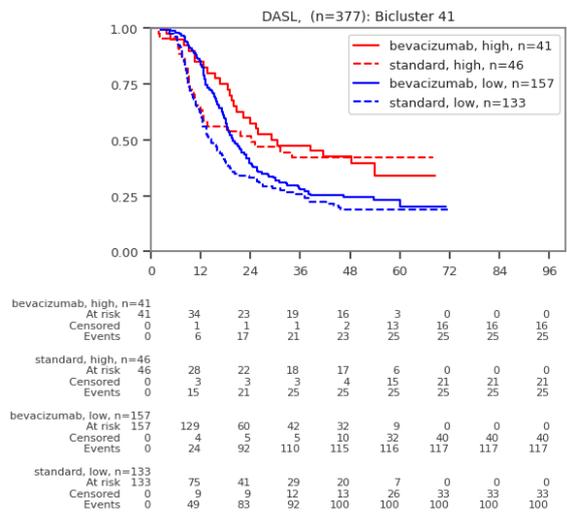

c

d

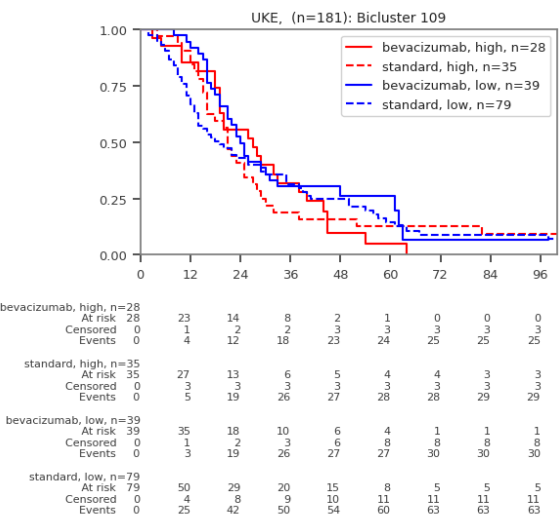

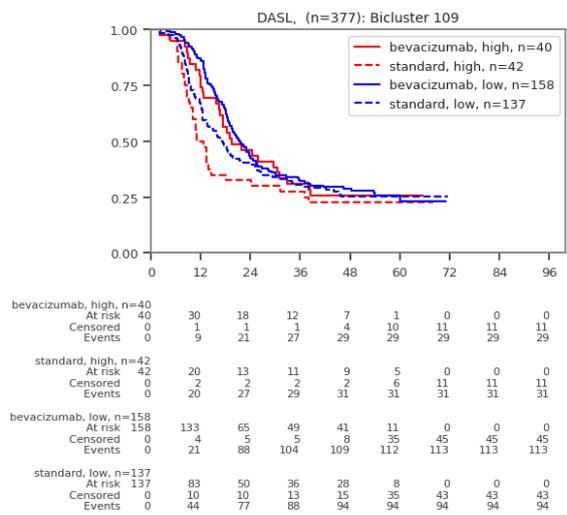

e

f

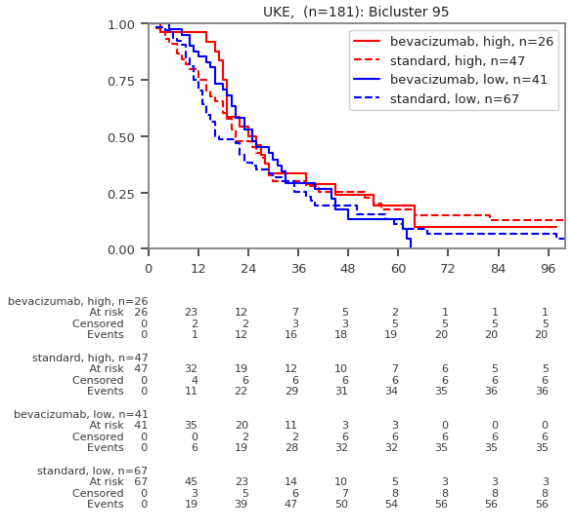
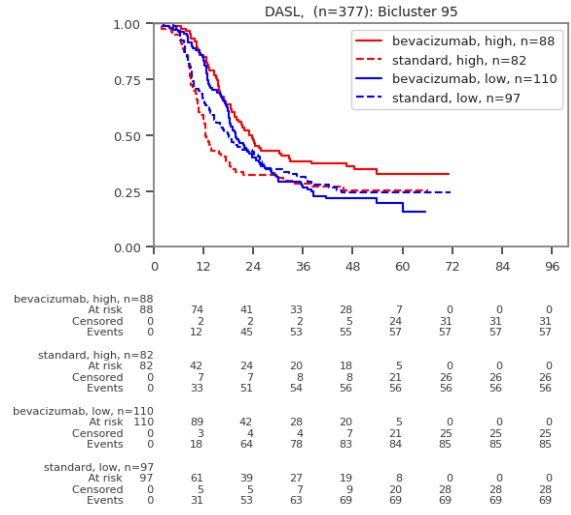

g  h

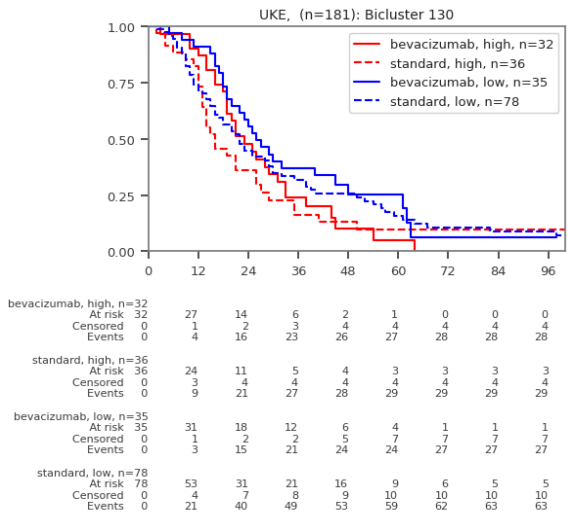
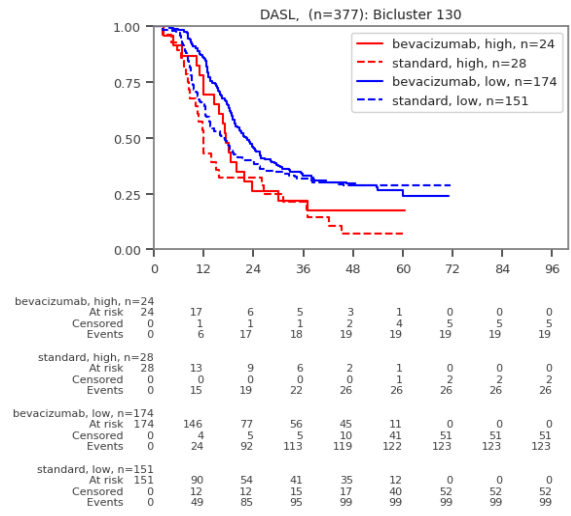

k  l

**Supplementary Figure S6.** Kaplan-Meier plots for PFS of patient subgroups of the UKE and DASL cohorts stratified by treatment (solid line - bevacizumab; dashed line - standard chemotherapy) and expression of genes comprising biclusters 70 (**a,b**), 41 (**c,d**), 109 (**e,f**), 95 (**g,h**), and 130 (**k,l**) prioritized by random survival forests as the best predictors of PFS.

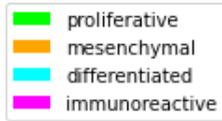
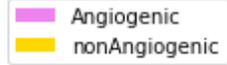
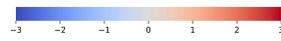
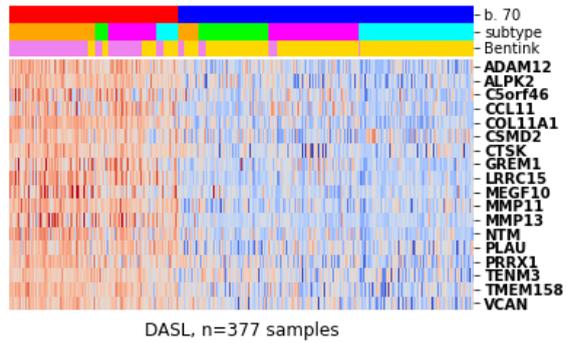
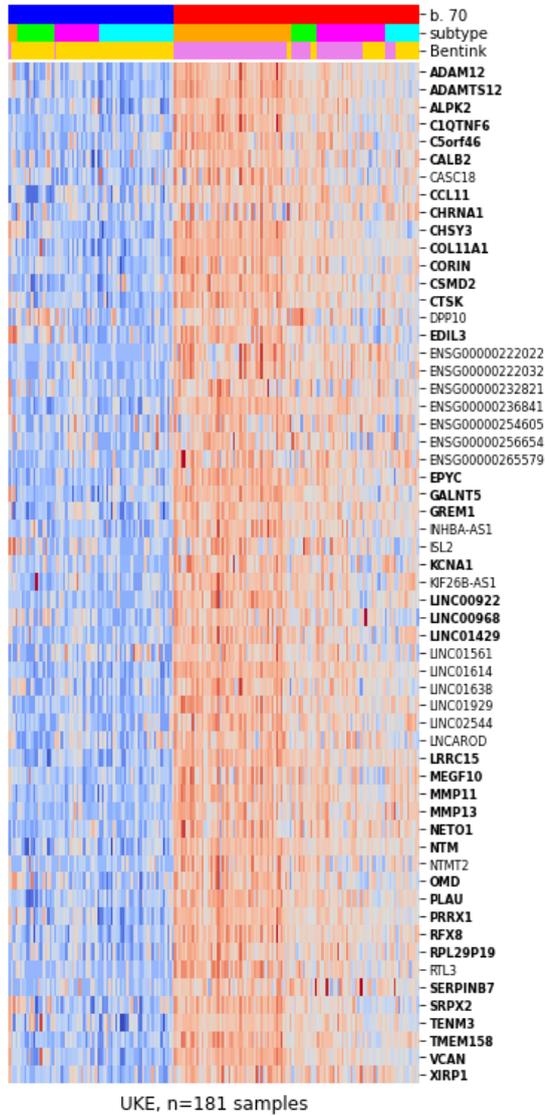
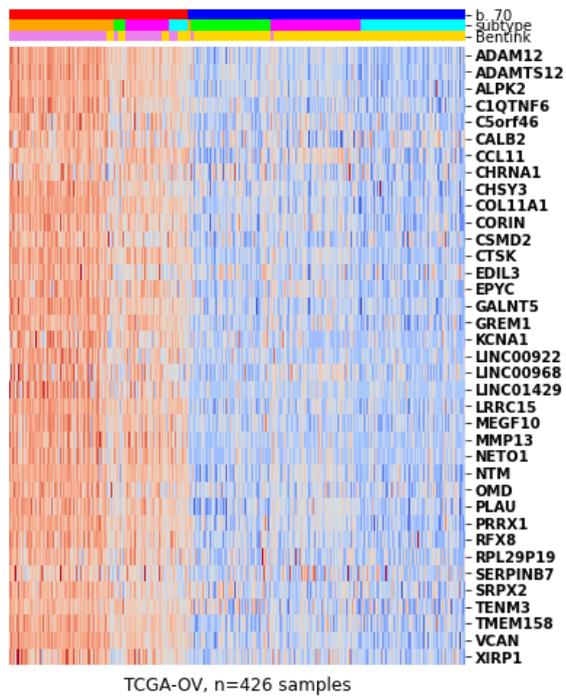

**Supplementary Figure S7**. Tumor stratification by Bicluster 70 is highly correlated with pro-/anti-angiogenic classification defined by Bentink et al. (pro-angiogenic - pink, anti-angiogenic - yellow). **a**. UKE. **b**. DASL **c**. TCGA-OV.

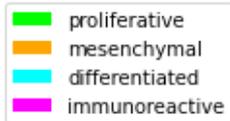
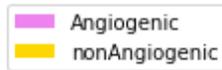
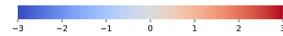
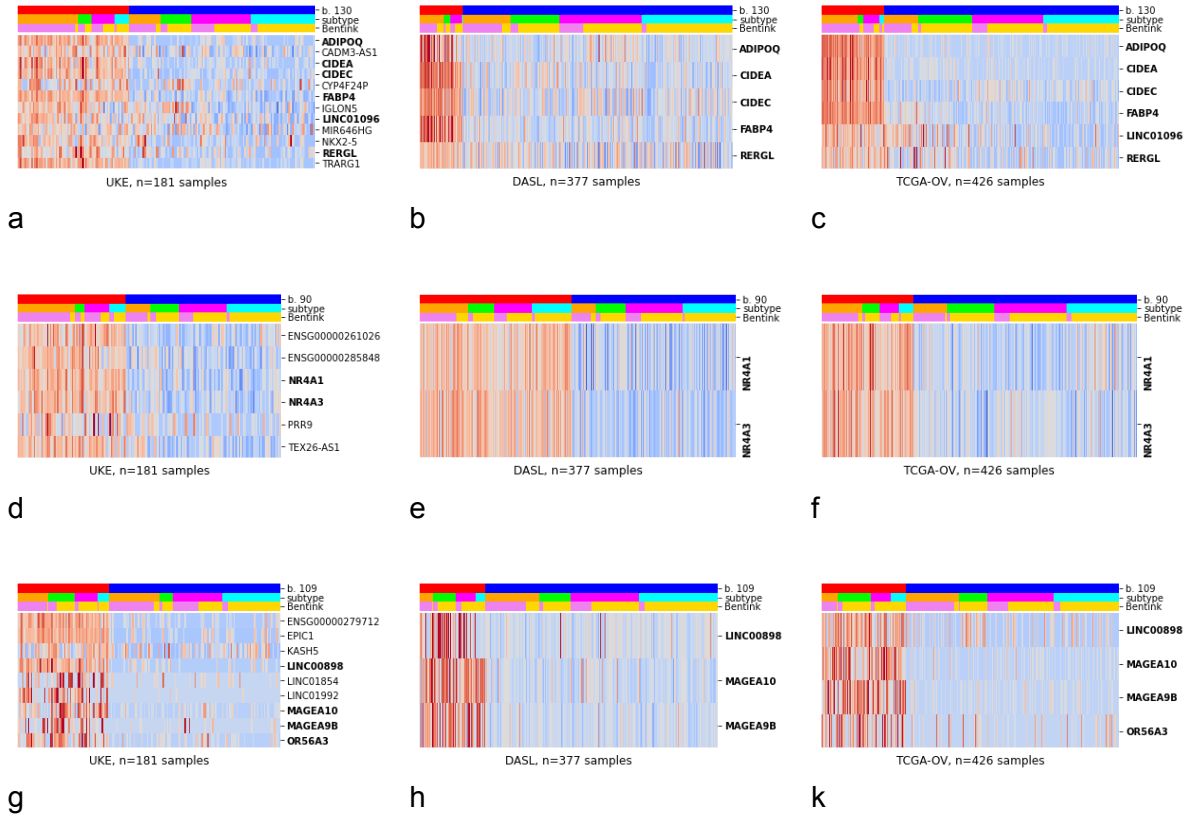

**Figure S8**. Replication of biclusters in the UKE (**a,d,g**), DASL (**b,e,h**) and TCGA-OV (**c,f,k**) datasets. **a-c**. Bicluster 130. **d-f**. Bicluster 90. **g-k**. Bicluster 109.

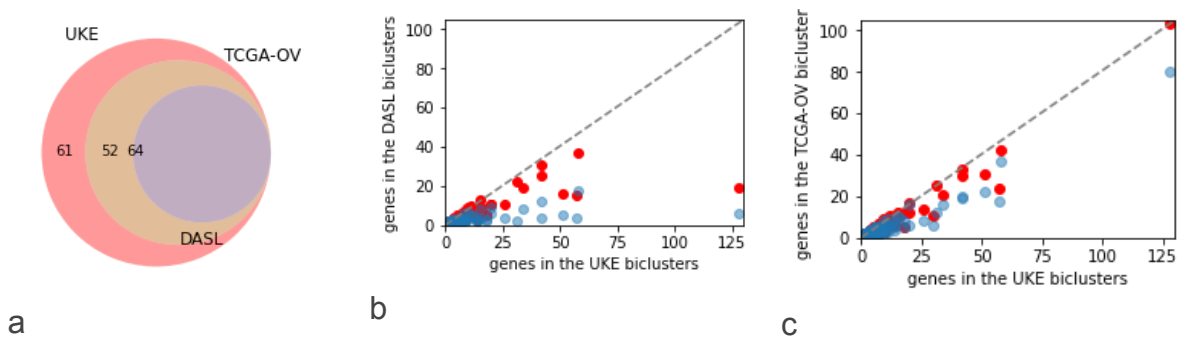

**Supplementary Figure S9.** Replication of biclusters identified in the UKE cohort in the TCGA-OV and in the DASL data. **a.** the number of biclusters replicated in each cohort. For each bicluster discovered in the UKE, in the validation cohort (TCGA-OV (**b**) or DASL(**c**)) were split by 2-means into two clusters based on the expression of bicluster genes. Genes with SNR<0.5 were removed and only biclusters consisting of at least two genes were kept. The number of UKE bicluster genes (x-axis) presented in the replication dataset is shown in red, and the number of genes retained in the bicluster is shown in blue.

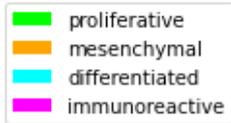
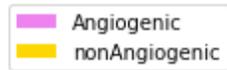
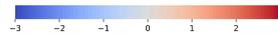
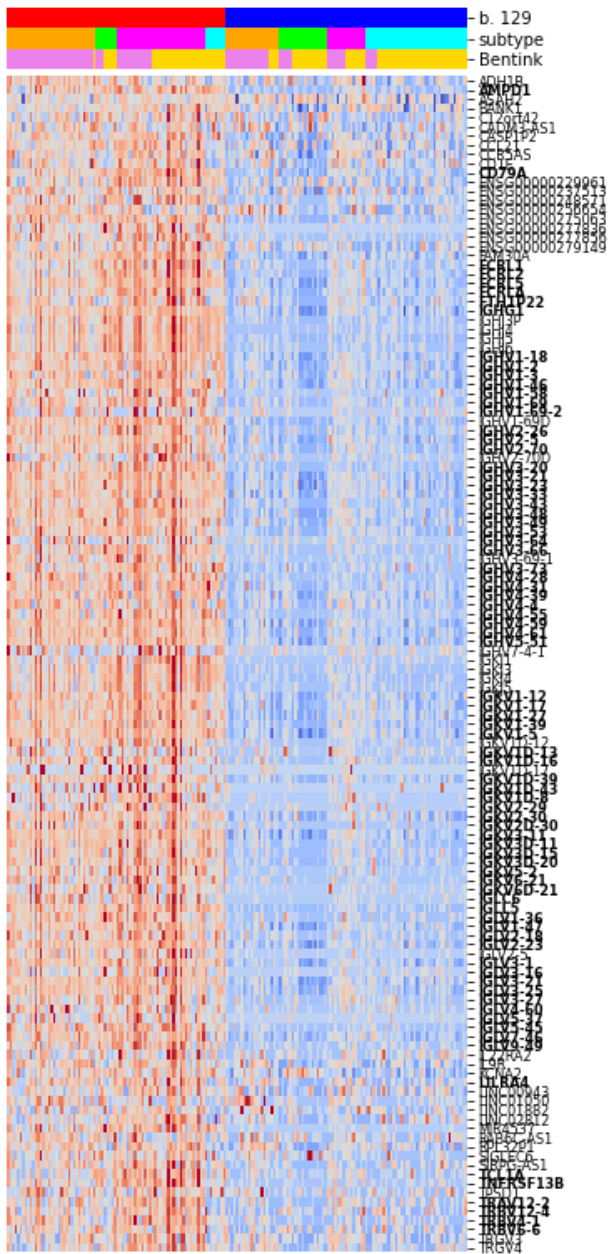
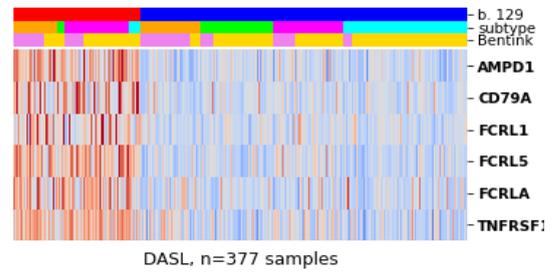
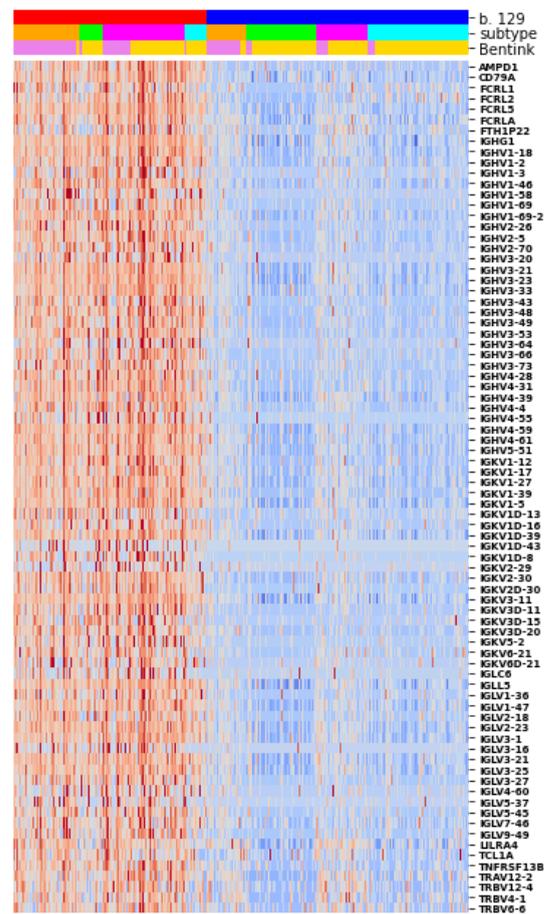

**Supplementary Figure S10.** Bicluster 129 in the UKE (**a**), DASL (**b**) and TCGA-OV (**c**) data.

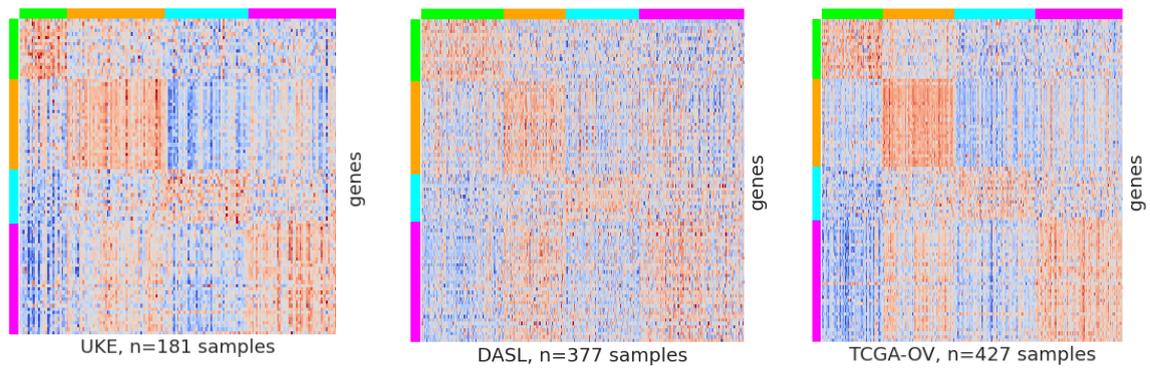

**Supplementary Figure S11.** Molecular subtypes of ovarian cancer predicted in the UKE (**A**), DASL (**B**), and TCGA-OV (**C**) cohorts assigned by the *consensusOV* classifier[2]. Normalized and standardized expressions of 98 genes with subtype-specific expressions manually curated from Verhaak et al.[3] and Talhouk et al.[4] are shown (see Methods for details).